\documentclass[rnote]{aa} 
\usepackage{graphicx}
\usepackage{txfonts}
\usepackage{natbib}
\usepackage[english]{babel}
\usepackage{lscape}
\bibpunct{(}{)}{;}{a}{}{,} 
\begin{document}
   \title{Variable stars in the globular cluster M28 (NGC~6626)}

   \author{G. Prieto,
          \inst{1,2}
		  M. Catelan,
          \inst{1,3}		  
		  R. Contreras Ramos,
		  \inst{1,4}
		  B. J. Pritzl, 
		  \inst{5}
		  H. A. Smith,
		  \inst{6}		  
		  \and 
		  J. Alonso-Garc\'{i}a
		  \inst{1,3}\fnmsep\thanks{Based on observations obtained with the SMARTS Consortium 1.3m telescope at the Cerro Tololo  Inter-American Observatory, Chile}		  
          }

   \institute{Pontificia Universidad Cat\'olica de Chile, Facultad de F\'{i}sica, 
       Departamento de Astronom\'\i a y Astrof\'\i sica, \\ Av. Vicu\~{n}a Mackenna 4860, 
       782-0436 Macul, Santiago, Chile
             \and
   Las Campanas Observatory, Carnegie Institution of Washington,
       Colina El Pino, Casilla 601, La Serena, Chile
	         \and
   The Milky Way Millennium Nucleus, Av. Vicu\~{n}a Mackenna 4860, 
       782-0436 Macul, Santiago, Chile		 
             \and
   Dipartimento di Astronomia, Universit\`a di Bologna, via Ranzani 1, 40127, Bologna, Italy
             \and
   Department of Physics \& Astronomy, University of Wisconsin-Oshkosh, Oshkosh, WI 54901, USA
             \and
   Dept.\ of Physics and Astronomy, Michigan State University, 
       East Lansing, MI 48824, USA}

   \date{A\&A, in press (June 20, 2012)}

 
  \abstract
   {We present a new search for variable stars in the Galactic globular cluster 
   M28 (NGC~6626).}
   {The search is based on a series of $BVI$ images obtained with the SMARTS
   Consortium's 1.3m telescope at Cerro Tololo Inter-American Observatory, Chile.}
   {The search was carried out using the ISIS v2.2 image subtraction package. 
   }
   {We find a total of 25 variable stars in the field of the cluster, 9 being new 
   discoveries. Of the newly found variables, 1 is an ab-type RR Lyrae star, 
   6 are c-type RR Lyrae, and 2 are long-period/semi-regular variables. V22, 
   previously classified as a type II Cepheid, appears as a bona-fide RRc in 
   our data. In turn, V20, previously classified as an ab-type RR Lyrae, could
   not be properly phased with any reasonable period.}
   {The properties of the ab-type RR Lyrae stars in M28 appear most consistent 
   with an Oosterhoff-intermediate classification, which is unusual for bona-fide
   Galactic globulars clusters. However, the cluster's c-type variables do not 
   clearly support such an Oosterhoff type, and a hybrid Oosterhoff I/II system 
   is accordingly another possibility, thus raising the intriguing possibility 
   of multiple populations being present in M28. Coordinates, periods, and light 
   curves in differential fluxes are provided for all the detected variables.}


   \authorrunning{Prieto et al.}
   \titlerunning{Variable Stars in M28}
			   
   \maketitle
%

\section{Introduction}\label{sec:intro} 

M28 (NGC~6266) is a moderately reddened [$E(B\!-\!V) = 0.40$~mag] and 
bright ($M_V = -8.16$~mag) globular cluster (GC) located at a low Galactic latitude 
($b = -5.58$).\footnote{Unless otherwise noted, all cluster parameters in this 
paper are from \citet[][Dec. 2010 update]{wh96}.}  
Though relatively close, at a distance from the Sun of only 5.5~kpc, 
it remains a relatively ill-studied cluster, likely due to the unfavorable  
(and highly variable) foreground reddening. 
Indeed, variability studies of the cluster have so far 
been mostly restricted to photographic data \citep{wb90,wc90}, 
and modern color-magnitude diagrams (CMDs) have 
only recently been presented in the literature 
\citep*{tdea96,area00,vtea01,jagea12}. 

Still, M28 appears as a particularly interesting object for at least three  
reasons: first, it has been found to have a disk-like orbit 
\citep{rc91,ch93}, thus making it one of the most metal-poor members of the so-called 
``thick disk'' family of GCs \citep[see, e.g.,][]{az88}. Second, among 
clusters of similar metallicity (${\rm [Fe/H]} \simeq -1.32$), M28 stands out as having 
a horizontal branch (HB) morphology strongly skewed towards the blue 
\citep[see also][]{ga81}, thus making it a ``second parameter'' cluster. 
The latter, if interpreted in terms of age, would accordingly indicate 
that the metal-poor tail of the thick disk is at least as old as the oldest components of 
the Galactic halo. Third, the cluster has been classified into an \citet{oo39,oo44} I 
(OoI) type, which is consistent with the relatively high metallicity of the cluster but 
in conflict with its HB type: with rare exceptions, blue HB clusters are commonly 
associated with type OoII \citep{rcea05,rcea10}. 

According to the \citet{ccea01} catalog,\footnote{\tiny\tt http://www.astro.utoronto.ca/$\sim$cclement/read.html} 
there are at present 24 variable stars known 
in the field of the cluster \citep[in addition to a millisecond pulsar;][]{alea87}. 
Of these, V1-V16 were discovered by \citet{hs49}, whereas 
V17-V24 were first discovered/reported on by \citet{wsh82,wsh84}. Updated ephemerides 
for these stars were provided in \citet{wb90} and \citet{wc90}, which remain the most recent 
papers to deal with the (photographic) light curves of M28 variable stars in a 
systematic way. \citet{rc91} reported, again based on photographic plates, 
on three additional candidate variables in the direction of the cluster, which however 
have not yet been systematically studied, or even incorporated into the electronic version 
of the \citeauthor{ccea01} catalog. 

There are strong reasons to believe that the current variable star tally for M28 is 
incomplete. First, while M28 is a very concentrated cluster, the photographic 
material which has been used in all previous variability studies does not allow the 
cluster core to be reliably resolved. Second, modern image-subtraction techniques
\citep[e.g.,][]{al98,ca00,db08}, 
when applied to modern CCD images, have recently been providing rich harvests of 
variable stars towards the centers of even the previously best studied globular 
clusters \citep*[see, e.g.,][for recent examples]{rcea10,ckea11}. 
Accordingly, the main purpose of this paper is to provide the first CCD-based 
time-series photometry for the central regions of M28, in order to search for 
additional variables that may have gone unnoticed in previous studies and to provided 
updated periods and light curves for previously studied variables. 

We begin in \S2 by describing out dataset and reduction techniques. In \S3, we 
describe our variability results. A summary of our results is finally provided in \S4.

\section{Observations and data reduction}\label{sec:obs} 
Our variability search is based on a total of 39 images in each of the Johnson-Cousins $BVI$ 
filters, obtained in service mode with the SMARTS 1.3m telescope at CTIO. The exposure times 
were 70, 25, and 15 seconds in $B$, $V$, and $I$, respectively. The images cover the period 
between June 2 and July 24, 2004, with 6 to 10 images obtained every 3 to 5 days. 

The images were reduced in the standard way, using IRAF tasks.\footnote{IRAF is distributed by the National Optical Astronomy Observatories, which are operated by the Association of Universities for Research in Astronomy, Inc., under cooperative agreement with the National Science Foundation.} A search for stellar variability was performed using ISIS v2.2 \citep{al98,ca00}, but modifying the search module so as to include candidates from all the available filters. For the variable star candidates detected, periods of variability were obtained using phase dispersion minimization \citep[PDM;][]{rs78}. 

 \begin{figure*}
   \centering
      \includegraphics[width=15cm]{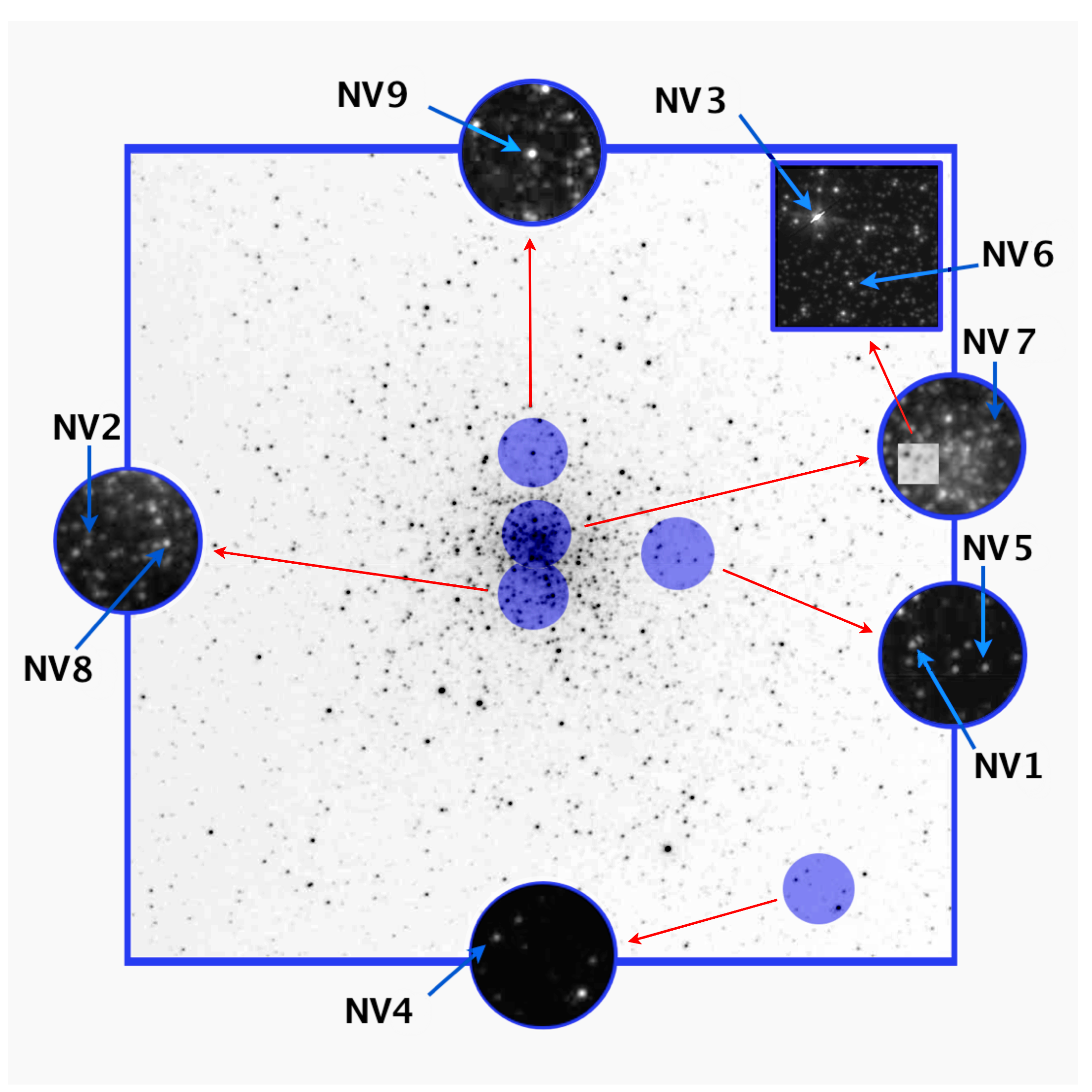}
   \caption{
   Finding chart for the newly discovered variable stars in M28. The large image
corresponds to a $V$-band image of the cluster, with the scale of the drawn blue
square being $6.5 \times 6.5 \,\, {\rm arcmin}^2$. North is up and East is to the left.
The five inner regions of the cluster where new variables have been detected are
shown as blue circles superimposed on the main image. Blow-ups are shown for each
of these regions, based on an F555W ($V$-band) {\em Hubble Space Telescope} image
obtained with the WFPC2 camera (ID 6625, PI R. Buonanno). Each such circle covers
a region of 30~arcsec diameter. In the top right, a $7.5 \times 7.5 \,\, {\rm arcsec}^2$ 
subregion of the circle containing NV3, NV6, and NV7 is shown on an expanded scale, for
clarity.   
   }
 \label{fig:chart}
 \end{figure*}

\section{Variable stars}\label{sec:vars}

Based on the variability search conducted with ISIS, we were able to find 25 variable stars with good photometry to reveal clean light curves. A finding chart with the new discoveries is given in Figure~\ref{fig:chart}. (For clarity, previously known variables are not shown in this chart. Finding charts indicating the positions of those variables, including those that fall outside our field of view, are provided by \citeauthor{wsh84} \citeyear{wsh84}.) Light curves for the individual variable stars (in relative flux units) are provided in Figures~\ref{fig:rrab} to \ref{fig:lpv} ({\em electronic version only}). The detected variables include 6 ab-type RR Lyrae, only one of which is a new discovery. In addition, 9 c-type RR Lyrae were found, 6 of which being new discoveries, of which one is a possible field interloper. The previously reported RR Lyrae stars are confirmed in our data, as are the 4 type II Cepheids. However, two of these candidate Cepheids could not be cleanly classified on the basis of our data. In particular, for V21 we are not able to phase the data properly, which might suggest that the star is a long-period variable (LPV). This is surprising, since \citet{wsh84} present a nicely phased light curve for this star, using a period of 29.93~d. Unfortunately, the star appears saturated in the \citet{jagea12} study; however, its position in our own preliminary CMD (not shown) is not inconsistent with an LPV classification. For V22, instead of the favored period of 0.99538~d in the 2009 version of the \citet{ccea01} catalog, which also comes from \citet{wsh84}, our data favor a much shorter period, of 0.323~d, and so this star may be better classified as a c-type RR Lyrae. The position of the star in the \citeauthor{jagea12} CMDs is fully consistent with a c-type classification. In \citet{wsh82} that reported the discovery of V22, a period of 0.498828~d was originally given, whereas in \citet{wsh84} a period of 0.498~d was given as an alternative, with a note also emphasizing that the star is blended in their images, which is supported by the large amount of scatter in their derived light curve. V21 and V22 are both fairly close to the center for such early photographic work.

The case of V20 is particularly intriguing. While previously a period of 0.49774~d was reported \citep{wsh84}, and the star classified as an ab-type RR Lyrae, we could not phase our data properly with this period. A period around 0.4948~d provides a better fit to our data, though with significant scatter. While this is consistent with \citeauthor{wsh84}'s comment that the star is blended, its mean magnitude in our preliminary photometry is not noticeably brighter than the magnitudes of other RR Lyrae stars in the cluster, and neither is it much redder than many of the other RR Lyrae. The light curve shape supports an ab-type classification, even though a cleaner light curve for a definitive assessment would be desirable. The star's position in the \citet{jagea12} CMDs is also fully consistent with an RRab classification. Unfortunately, our dataset is insufficient for an assessment of the incidence of the \citet{sb07} effect among our studied stars. 

As to the remaining stars, namely V2, V3, V7, and V10, we confirm the results of previous studies, in that the stars appear to vary on long timescales. For V2 and V3, we provide periods for the first time. V7 had a previously reported period of 320~d, which is much too long a timescale for our dataset to provide meaningful information regarding periodicity. In any case, its position in the \citet{jagea12} CMDs clearly confirm that it is a fairly metal-rich red giant star in the background. Finally, for V10 we are again unable to phase the light curve with a significant period. The remaining known variables not in our study generally fall outside our field of view: V1, V8, V9 (all ab-type RR Lyrae), V6 (an LPV), V15, V16 (both ab-type RR Lyrae belonging to the field). V14 is likely not variable, according to the \citet{ccea01} catalog \citep[see][]{wb90}, even though \citet{wsh84} provide possible periods around 0.2688~d for this star. 

Most of the variable star candidates listed by \citet{rc91} could not be confirmed in our study. However, based on the position in their finding chart, and as indicated in Table~\ref{tab:var}, RC133 may be the same star as our NV5, even though the coordinates published in their paper do not exactly match ours. The matching to \citeauthor{rc91} was done by triangulation and by eye.

Table~\ref{tab:var} summarizes our results. Column 1 gives the star name, whereas columns 2 and 3 give their right ascension and declination, respectively (epoch J2000). We derived celestial coordinates for all stars detected in our photometric study by comparison with bright stars obtained from the Two Micron All Sky Survey \citep[2MASS;][]{msea06} catalog stars available through the Infrared Processing and Analysis Center (IPAC) website. Around 1300 stars were available as astrometric references within our image. A third-order polynomial fit, done with the IRAF task {\it mscpeak}, produced dispersions of ${\sigma}\sim0.2''$, consistent with the catalog precision. Column 4 gives our best-fitting period, with as many decimals places as justified on the basis of our data only, whereas column 5 gives the previously favored period. Column 6 gives the variability type favored by our data, whereas column 7 gives the previously favored variability type. The final column gives an alternative identification for the star, if available. Light curves for the individual variable stars (in relative flux units) are provided in Figures~\ref{fig:rrab} to \ref{fig:lpv} ({\em electronic version only}). 

Among the ab-type RR Lyrae in M28, based on the magnitude values published in the online \citet{ccea01} catalog, we associate as likely field variables not only stars V15, V16, and V24 (all of which are background stars), but also V9, which is about 1~mag brighter than the other bona-fide cluster RR Lyrae stars, and is thus likely a foreground RR Lyrae. The newly discovered RRab, NV1, appears to be a bona-fide cluster member, judging from its position in the \citet{jagea12} CMDs. 

As far as c-type RR Lyrae are concerned, no previous field interlopers had been known. Among the newly detected RRc's, only NV3 appears clearly to be a foreground RR Lyrae, being as it is about 2~mag brighter than other RR Lyrae stars in the cluster, again according to the \citet{jagea12} CMDs.  

For the RR Lyrae stars that are clearly cluster members, one obtains the following average quantities and population ratios: $\langle P_{\rm ab}\rangle = 0.602$~d (or 0.614~d, if we drop V20); $\langle P_{\rm c}\rangle = 0.304$~d; $f_{\rm c} = N_{\rm c}/N_{\rm c + ab} = 0.44$ (or 0.47, if V20 is excluded). In addition, the minimum ab-type RR Lyrae period ($P_{\rm ab,min}$) and maximum c-type RR Lyrae period ($P_{\rm c,max}$) are 0.4916~d and 0.329~d, respectively. These values are quite unusual among known GCs. In particular, both $\langle P_{\rm ab}\rangle$ and $P_{\rm ab,min}$ are more  typical of Oosterhoff-intermediate systems \citep[e.g.,][]{mc09,mcea12}. As discussed by \citet{mcea12}, these are the two quantities most strongly defining Oosterhoff status. Still, M28's $\langle P_{\rm c}\rangle$ value is clearly more typical of Oosterhoff type I systems, whereas the high c-type number fraction is more common in Oosterhoff type II systems. A ``hybrid'' Oosterhoff I/II system is accordingly another possibility, which raises the question of whether there might be multiple populations in this fairly massive cluster. Given the known connections between Oosterhoff status and formation history \citep[e.g.,][]{mc09,gc10,vb11,mcea12}, it would certainly be of interest to study the variable star content in M28 using a more extensive dataset.  

To close, we comment on the status of NV8. The star's position in the \citet{jagea12} CMDs, fairly close to the tip of the red giant branch and at a fairly red color, is consistent with an LPV classification. This is why this star is grouped with other LPVs in Figure~\ref{fig:lpv}. On the other hand, according to our time-series data, a period around 13.53~d provides a decent match of the light curve (Fig.~\ref{fig:lpv}). Such a period is, however, more consistent with a W Virginis (type II Cepheid) classification. Further analysis will clearly be necessary, before the variability nature of this star can be conclusively established.

\begin{table*}\label{tab:var}
\caption{Variable stars in M28}
\centering
\begin{tabular}{cccccccc}
\hline\hline
Name & RA (J2000) & DEC & Period & Old period & Type & Old type & Other ID \\
 & (hh:mm:ss.ss) & (dd:mm:ss.s) & (days) & (days) &  & \\ 
\hline
     V2   & 18:24:29.15 & -24:51:06.6& 58     &           & LPV 		& Irr       		\\
	 V3   & 18:24:30.17 & -24:50:18.5& 49 	  &           & LPV 		& Irr				\\
	 V4   & 18:24:30.12 & -24:51:36.4& 13.37  & 13.462    & W~Vir 	    & W~Vir				\\
	 V5   & 18:24:29.48 & -24:51:52.6& 0.645  & 0.644360  & RRab 		& RR0				\\
	 V7   & 18:24:44.19 & -24:50:30.6&        & 320.0     & LPV, f	    & Mira, f			\\
	 V10  & 18:24:39.51 & -24:53:26.1&        &  		  & LPV/Irr 	& Irr				\\
	 V11  & 18:24:31.48 & -24:51:33.0& 0.542  & 0.542767  & RRab 		& RR0	    		\\
	 V12  & 18:24:43.31 & -24:52:56.1& 0.578  & 0.578228  & RRab 		& RR0				\\
	 V13  & 18:24:25.77 & -24:52:33.6& 0.656  & 0.654923  & RRab, f? 	& RR0				\\
	 V17  & 18:24:35.84 & -24:53:15.8& 62 	  & 91.7 	  & RV~Tau	    & RV~Tau			\\
	 V18  & 18:24:36.54 & -24:51:49.9& 0.640  & 0.640151  & RRab 		& RR0				\\
	 V19  & 18:24:30.84 & -24:51:56.0& 0.3119 & 0.335     & RRc 		& RR1				\\
	 V20  & 18:24:33.11 & -24:51:42.9& 0.4948 & 0.49774   & RRab?		& RR0				\\
	 V21  & 18:24:32.94 & -24:51:59.0&        & 29.93     & LPV		    & W~Vir				\\
	 V22  & 18:24:30.98 & -24:52:01.8& 0.323  & 0.99538   & RRc 		& BL~Her			\\
	 V23  & 18:24:30.25 & -24:52:03.0& 0.292  & 0.29231   & RRc 		& RR1				\\
	 NV1  & 18:24:28.89 & -24:52:09.9& 0.748  &  		  & RRab 		& 					\\
	 NV2  & 18:24:33.59 & -24:52:32.3& 0.297  &  		  & RRc  		& 					\\
	 NV3  & 18:24:32.31 & -24:52:32.2& 0.323  &  		  & RRc, f  	& 					\\
	 NV4  & 18:24:24.13 & -24:54:36.6& 0.261  &  		  & RRc  		& 					\\
	 NV5  & 18:24:27.67 & -24:52:15.0& 0.311  &  		  & RRc  		& 		  & RC133?  \\
	 NV6  & 18:24:33.45 & -24:52:10.6& 0.308  &  		  & RRc  		& 					\\
	 NV7  & 18:24:32.17 & -24:52:00.4& 0.329  &  		  & RRc  		& 					\\
	 NV8  & 18:24:33.60 & -24:52:07.1&        & 13.53 	  & W Vir? LPV? & 					\\
	 NV9  & 18:24:33.05 & -24:51:28.9&        &  		  & LPV   		& 					\\
\hline
\end{tabular}
\end{table*}


\onlfig{2}{
 \begin{figure*}
   \centering
\includegraphics[width=.325\textwidth]{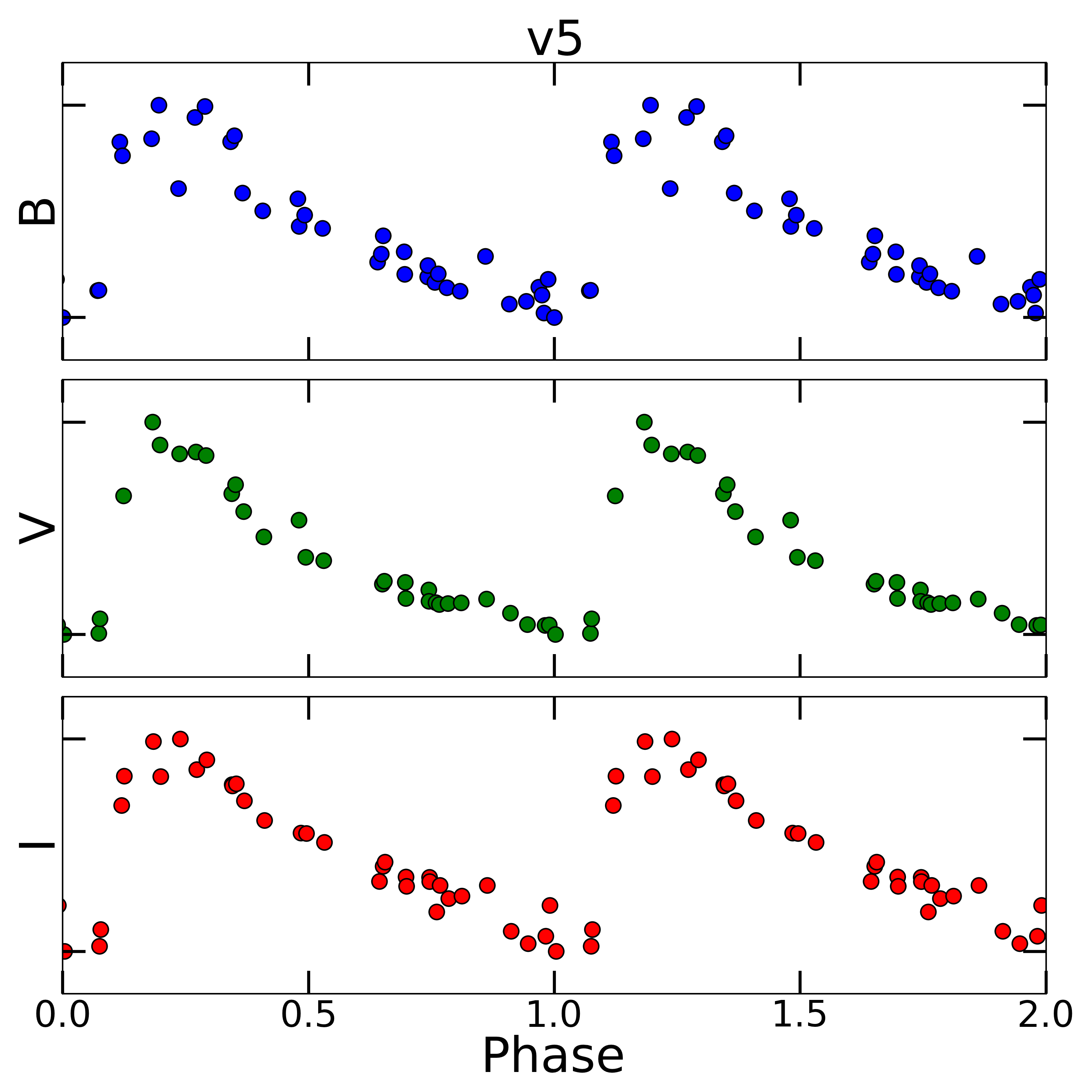}
\includegraphics[width=.325\textwidth]{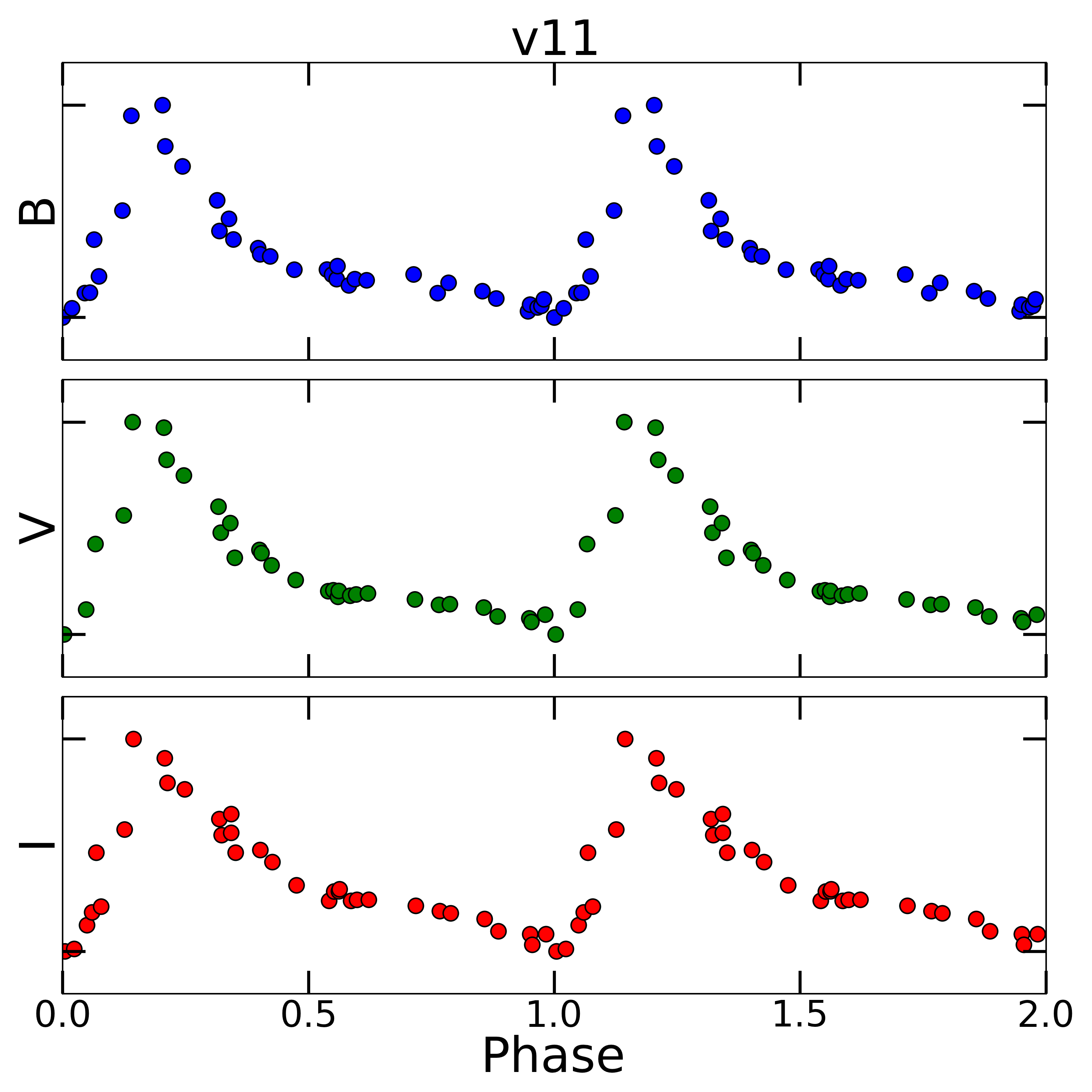}
\includegraphics[width=.325\textwidth]{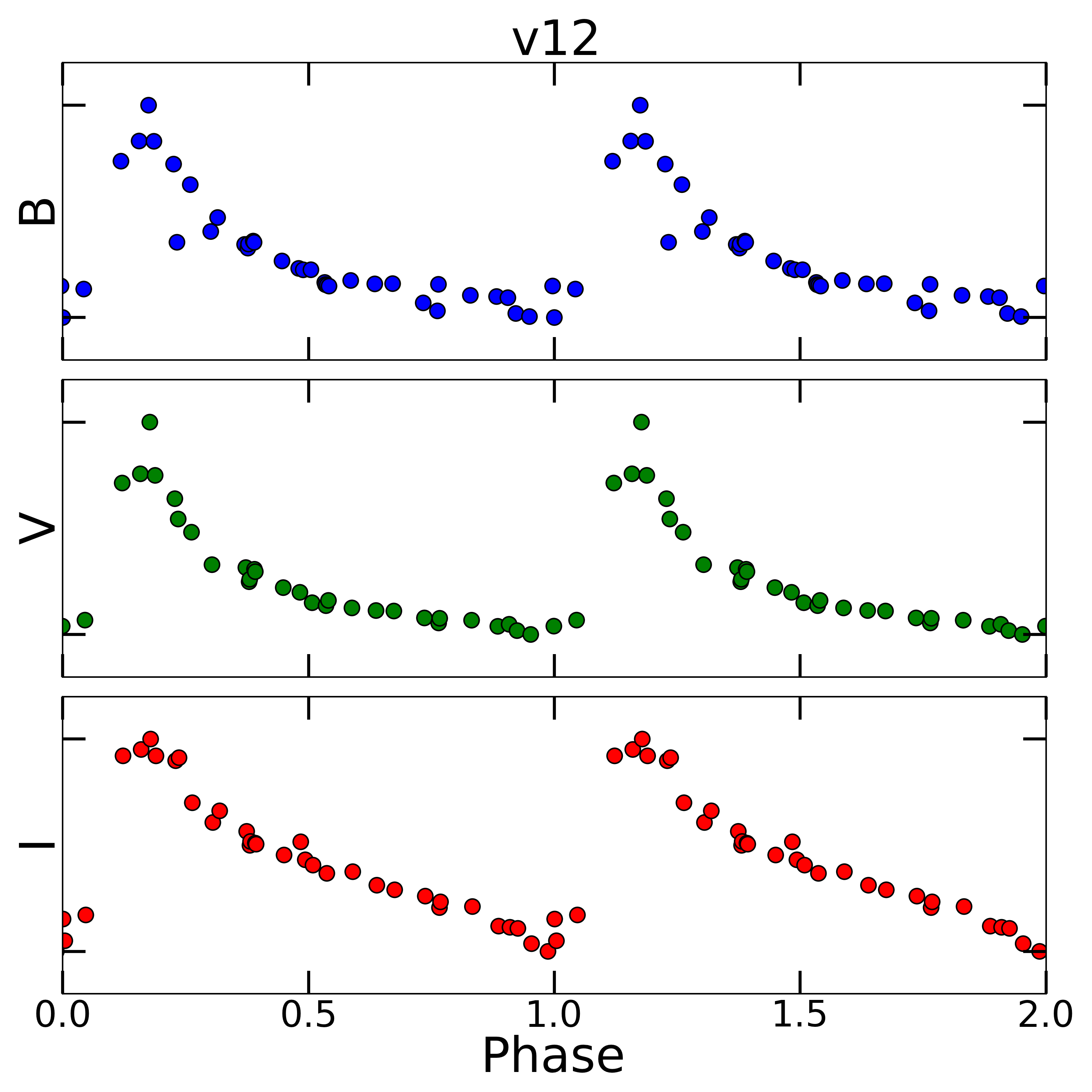}
\includegraphics[width=.325\textwidth]{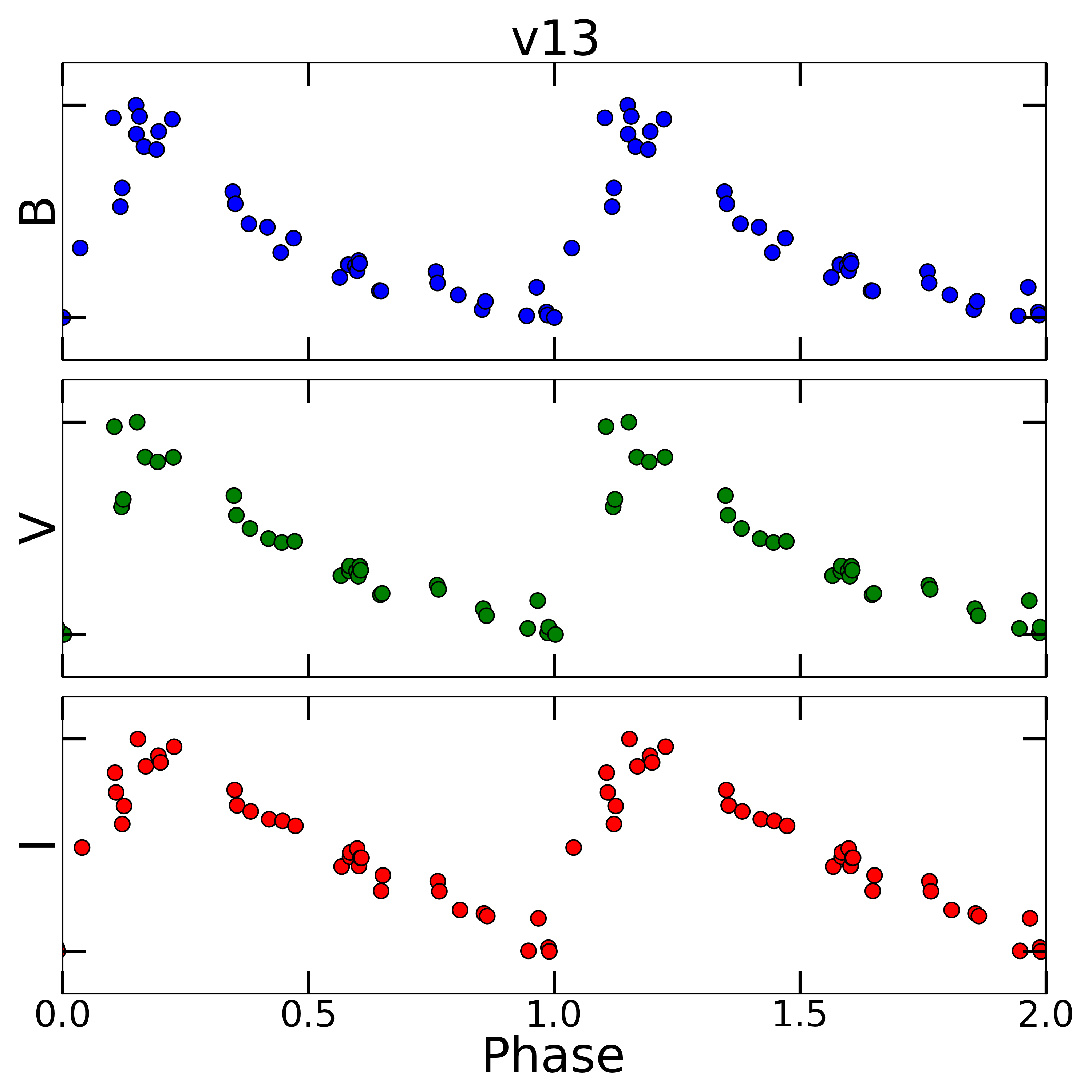}
\includegraphics[width=.325\textwidth]{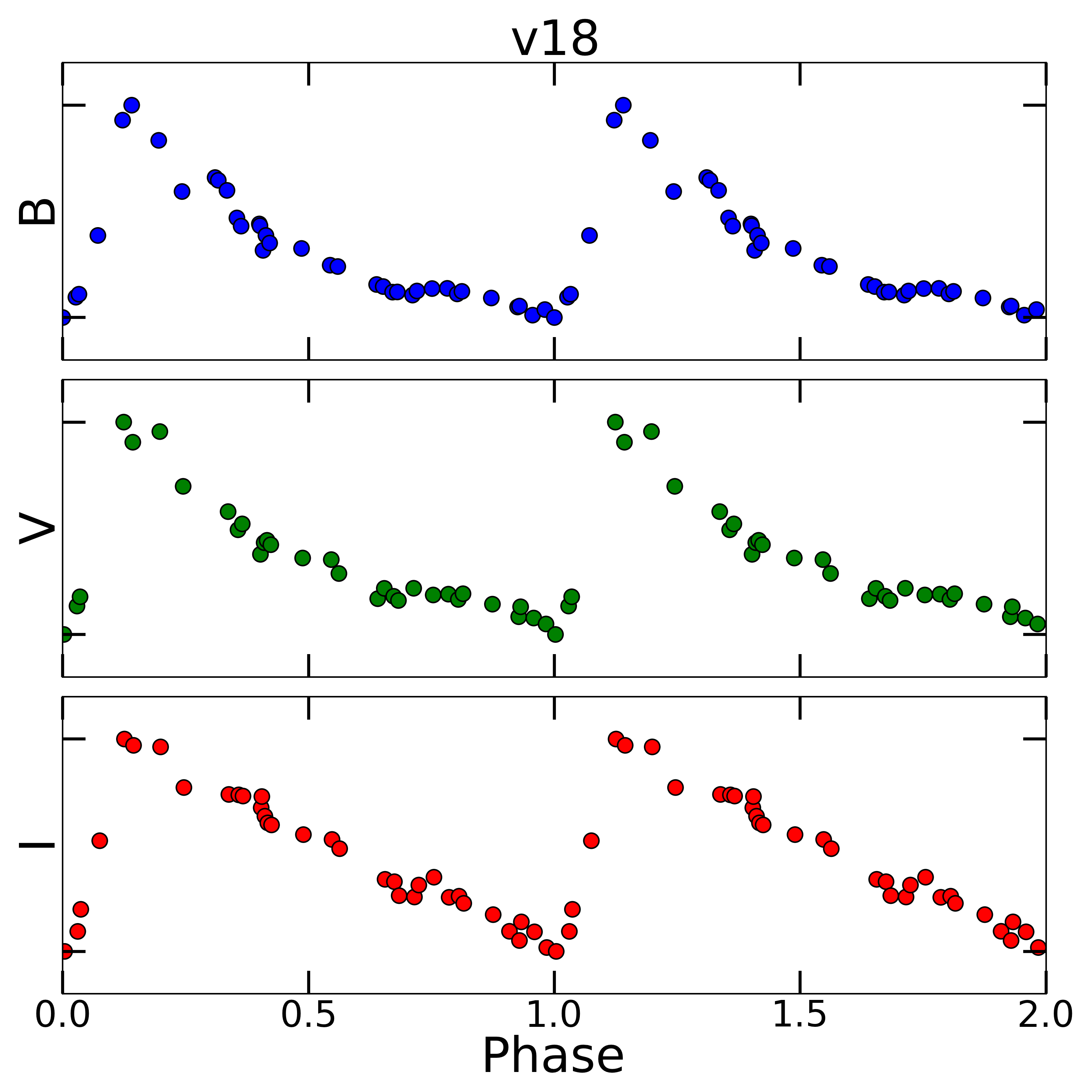}
\includegraphics[width=.325\textwidth]{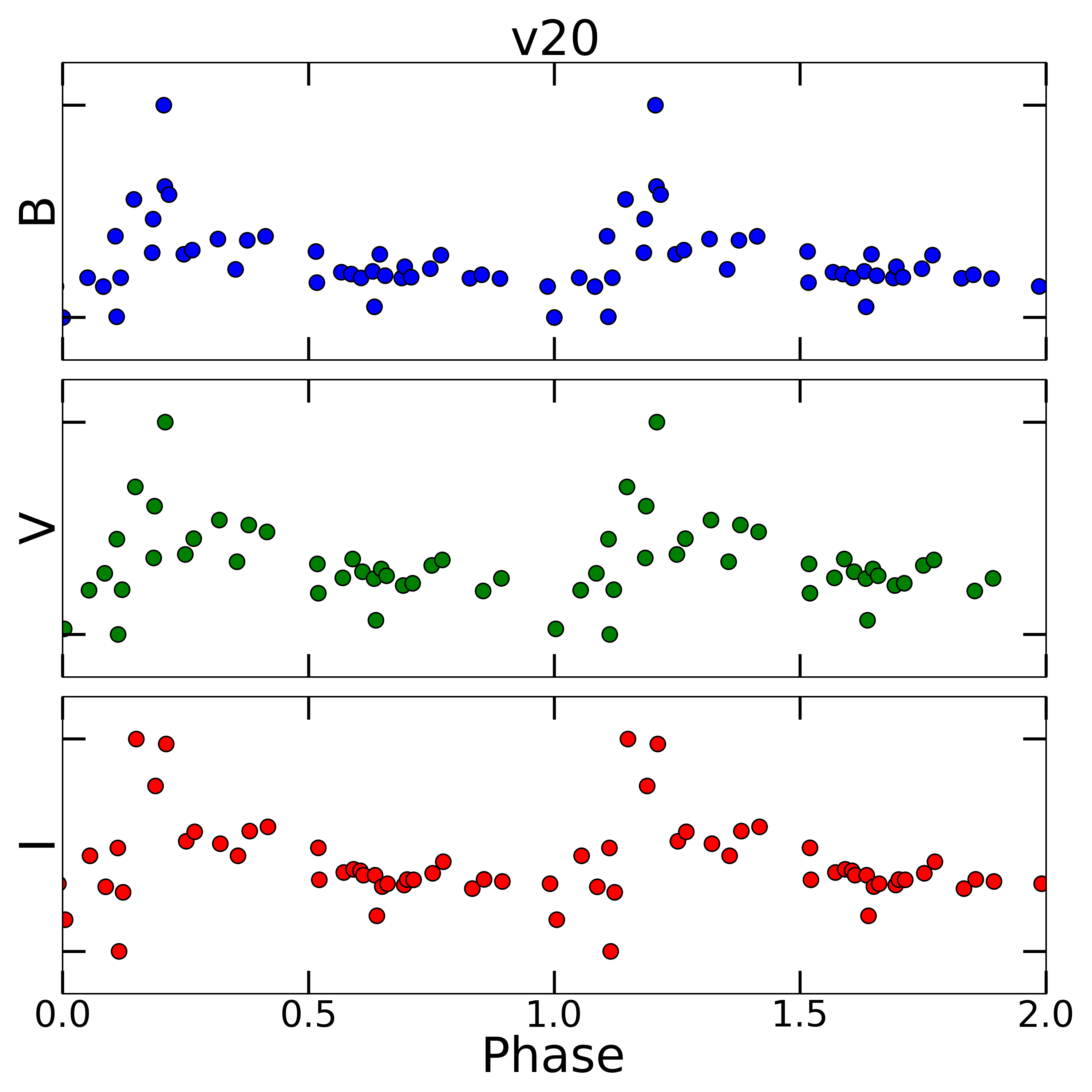}
\includegraphics[width=.325\textwidth]{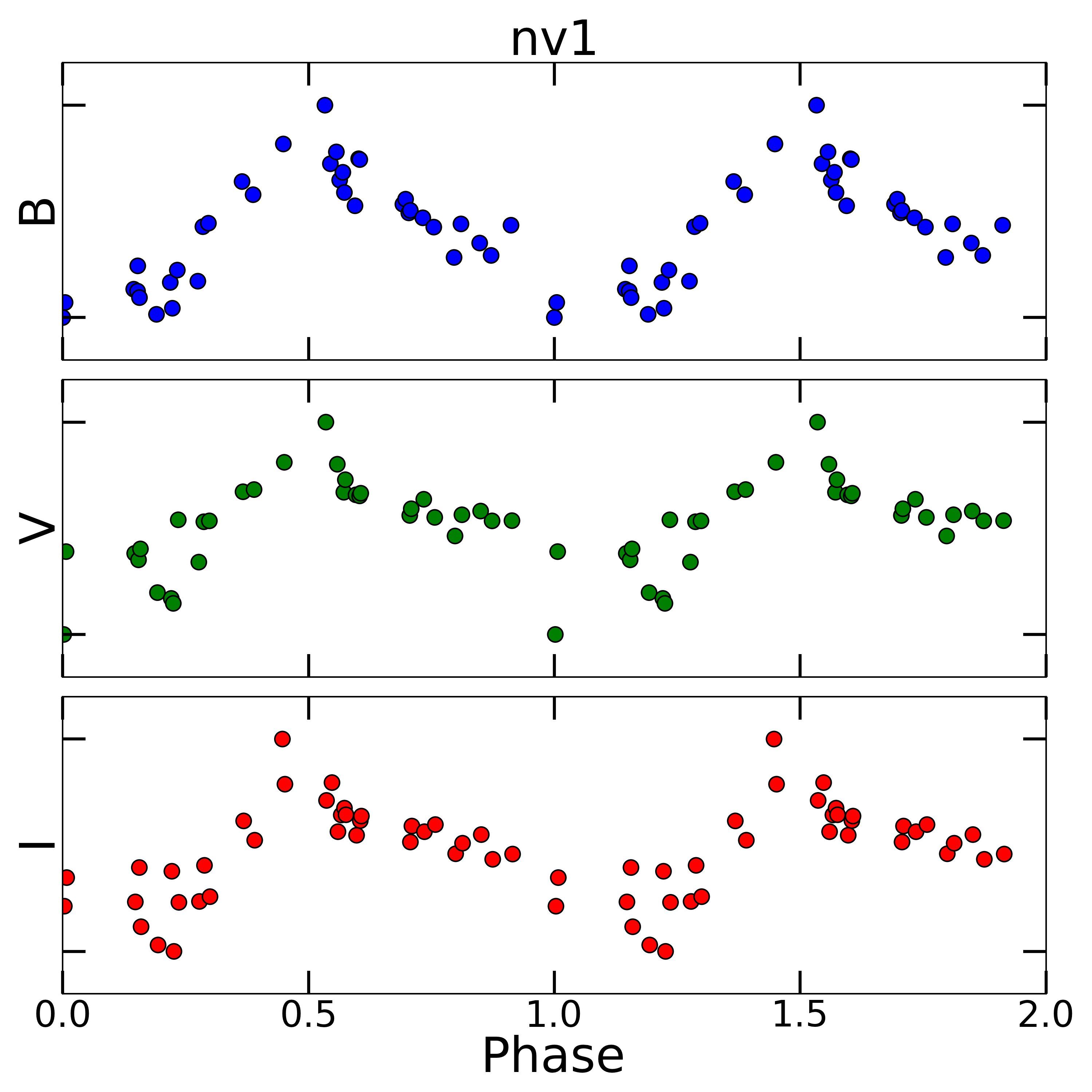}
   \caption{Light curves of the ab-type RR Lyrae detected in this work, in the 
    $B$ ({\em top panels}), $V$ ({\em middle panels}), and $I$ ({\em bottom 
	panels}), respectively.}
 \label{fig:rrab}
 \end{figure*}
}

\onlfig{3}{
 \begin{figure*}
   \centering
\includegraphics[width=.325\textwidth]{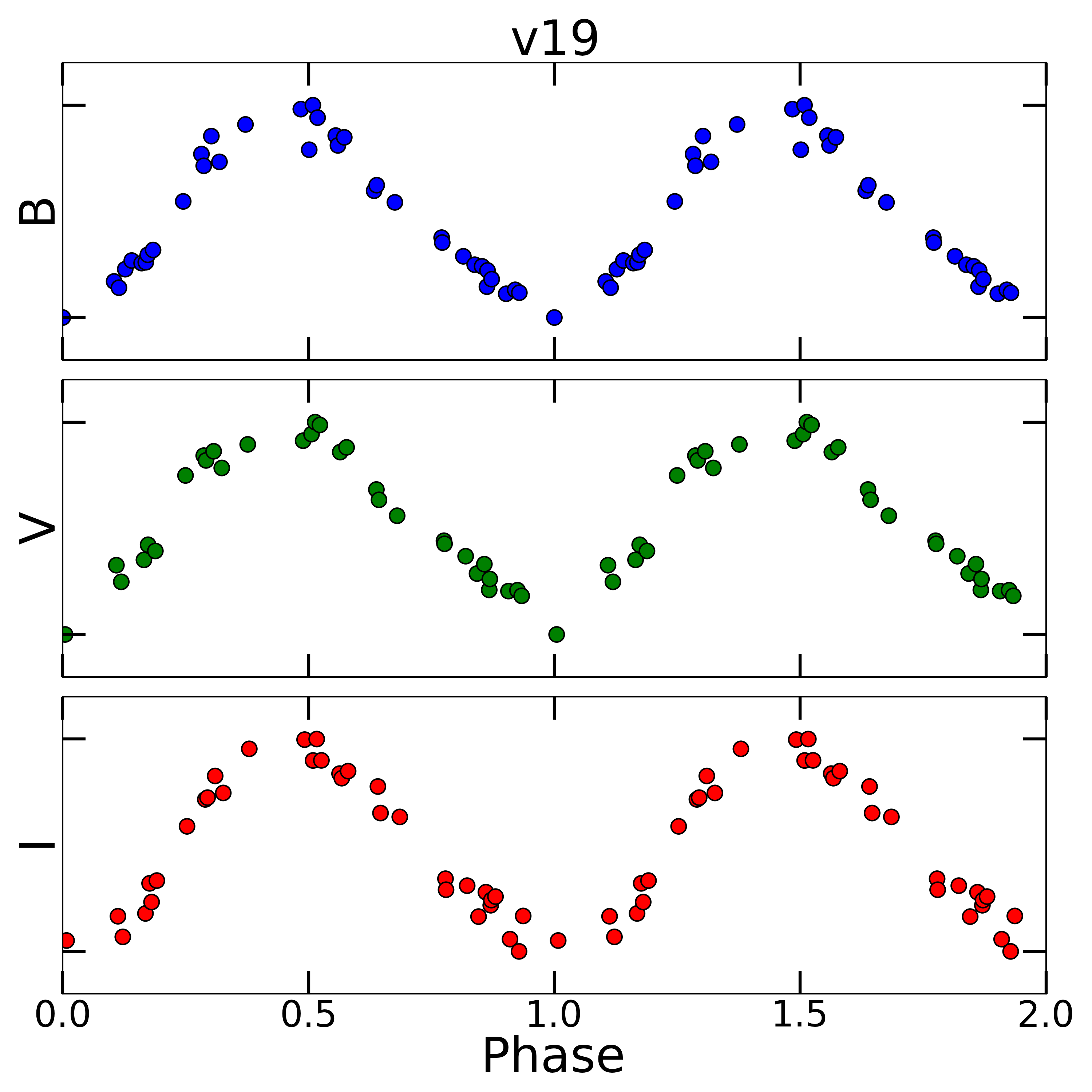}
\includegraphics[width=.325\textwidth]{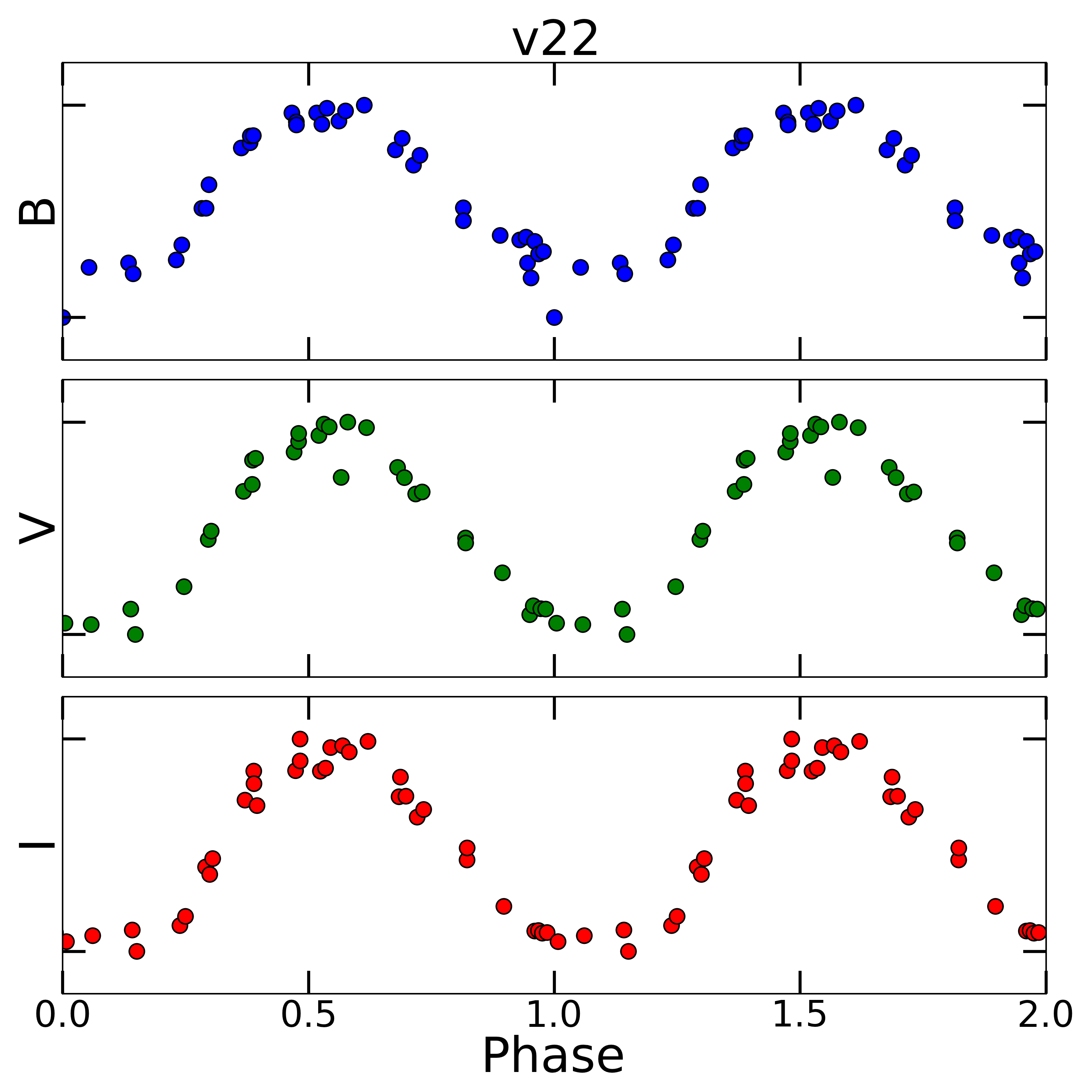}
\includegraphics[width=.325\textwidth]{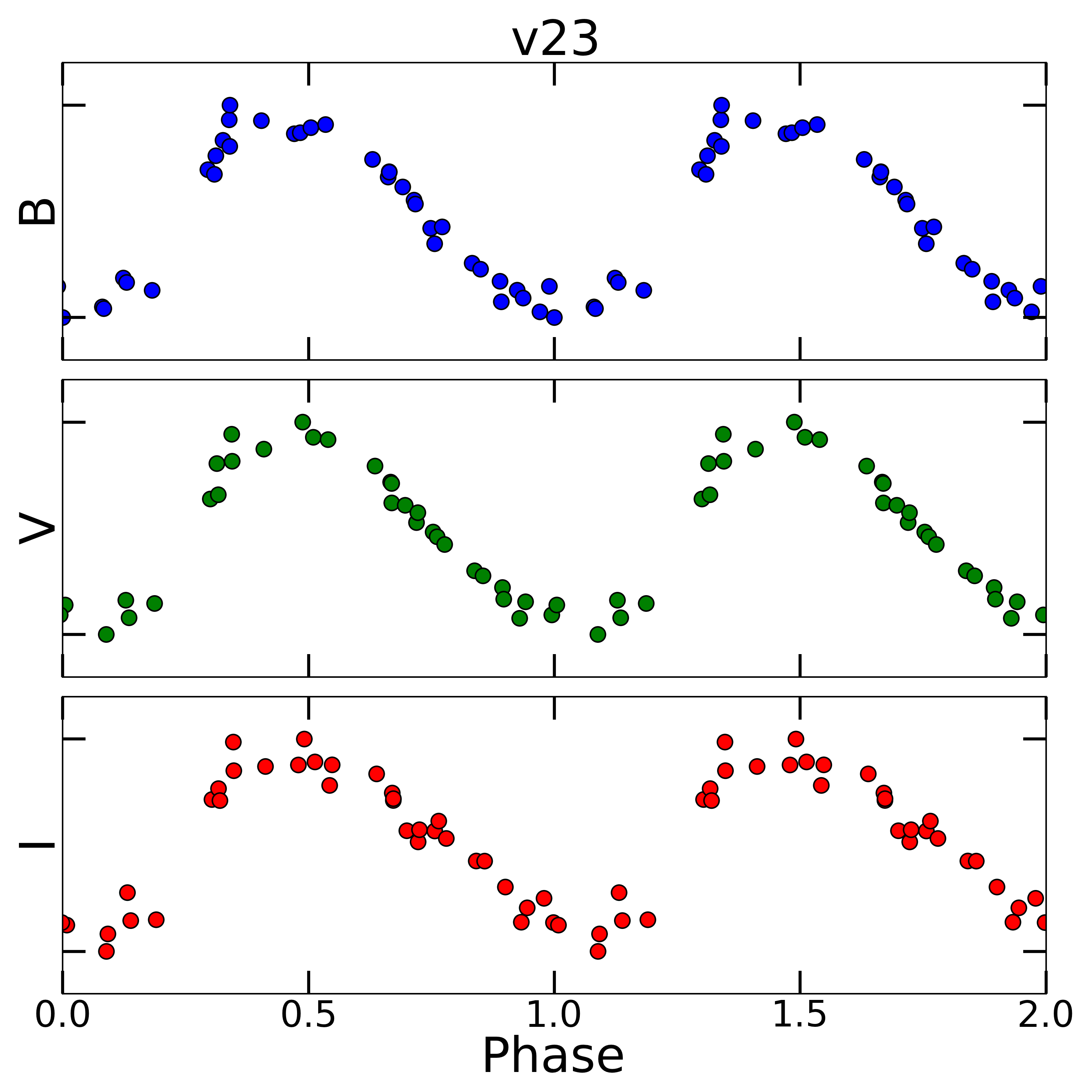}
\includegraphics[width=.325\textwidth]{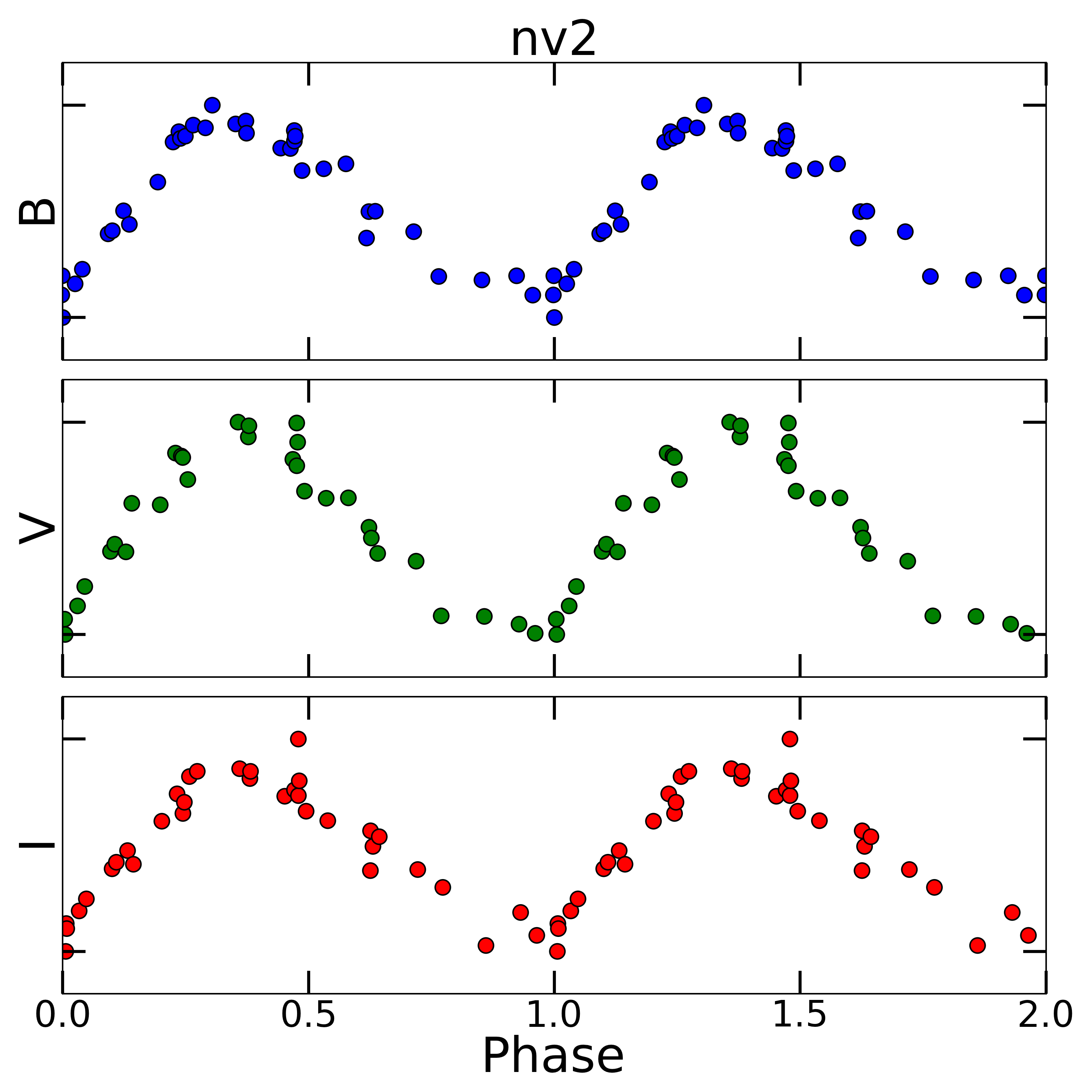}
\includegraphics[width=.325\textwidth]{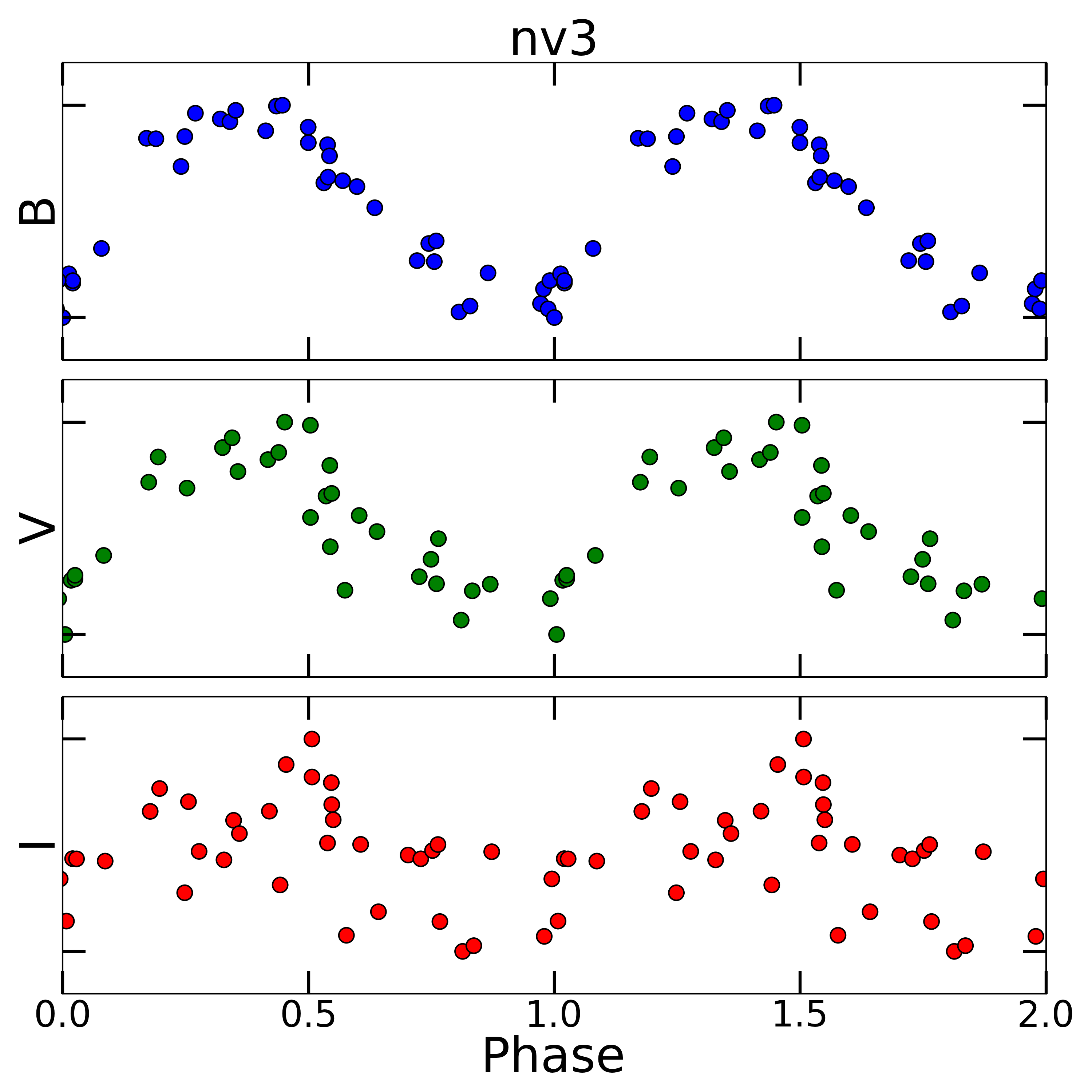}
\includegraphics[width=.325\textwidth]{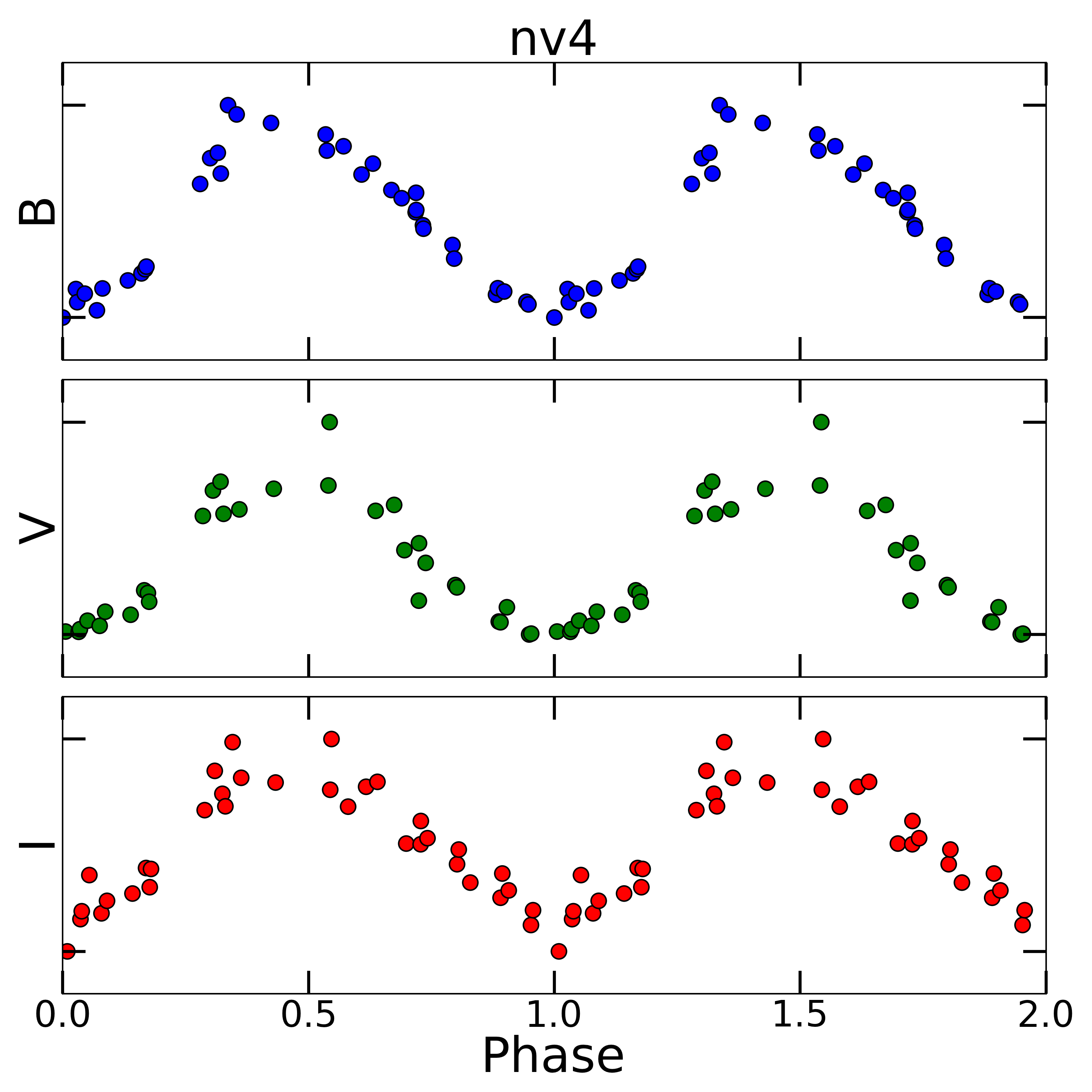}
\includegraphics[width=.325\textwidth]{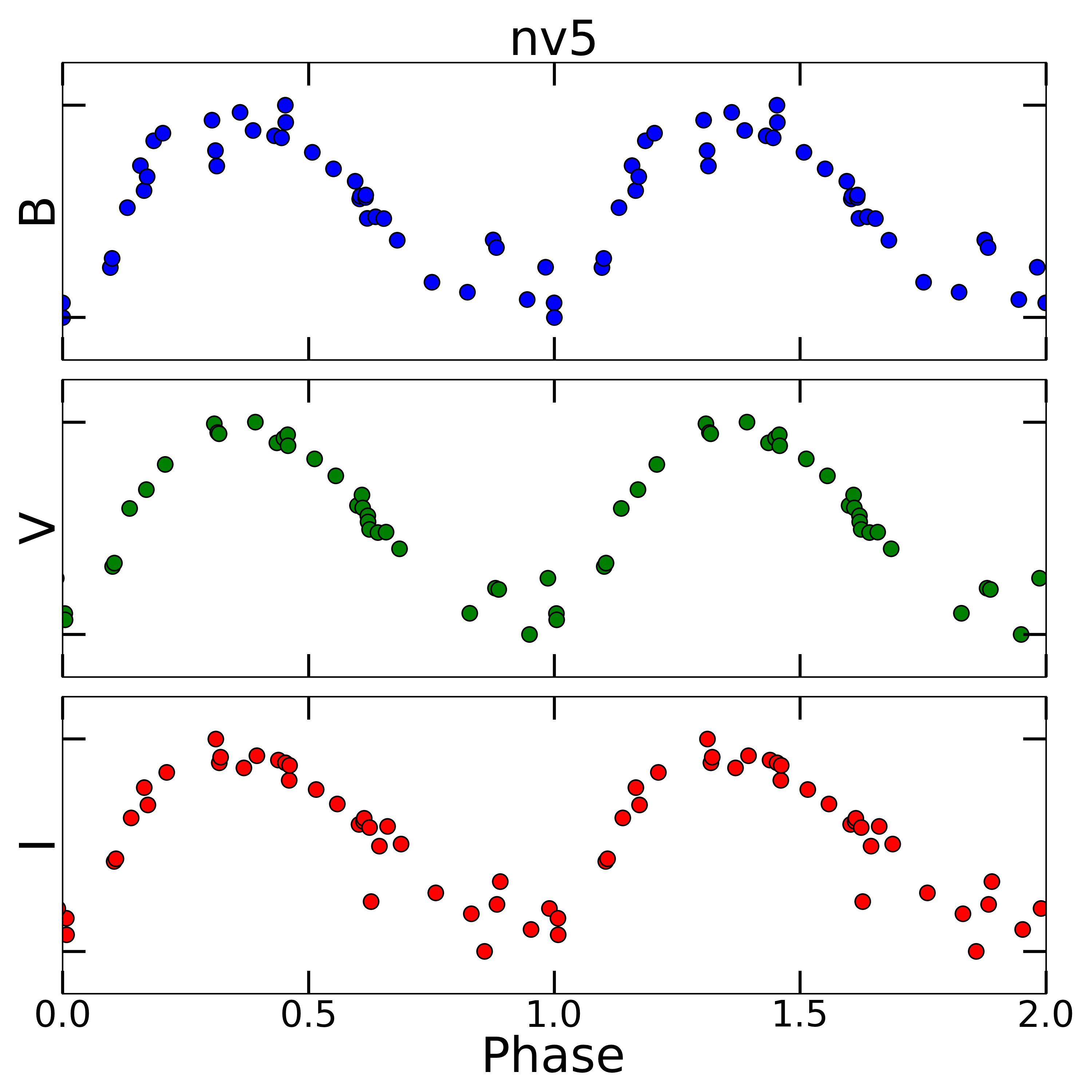}
\includegraphics[width=.325\textwidth]{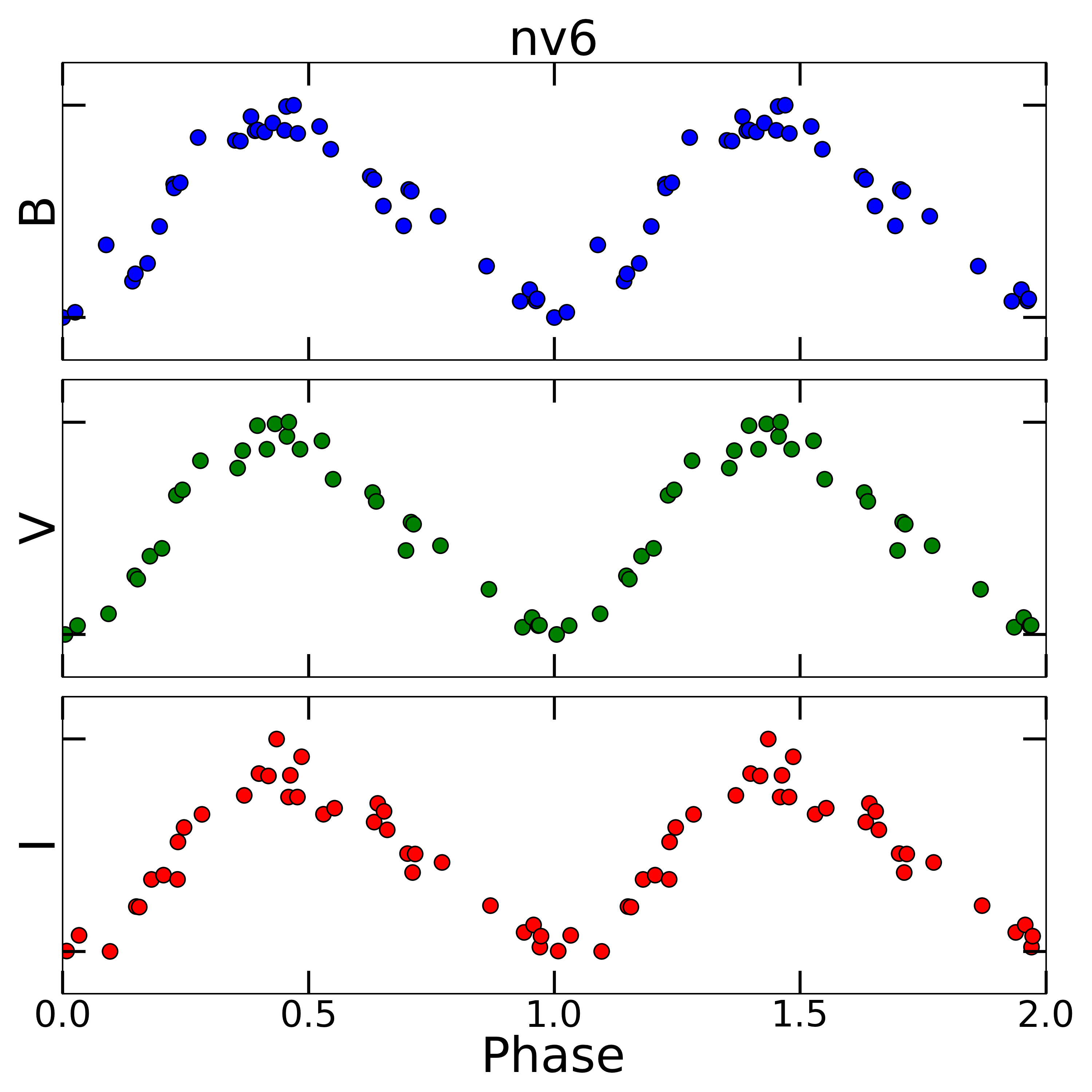}
\includegraphics[width=.325\textwidth]{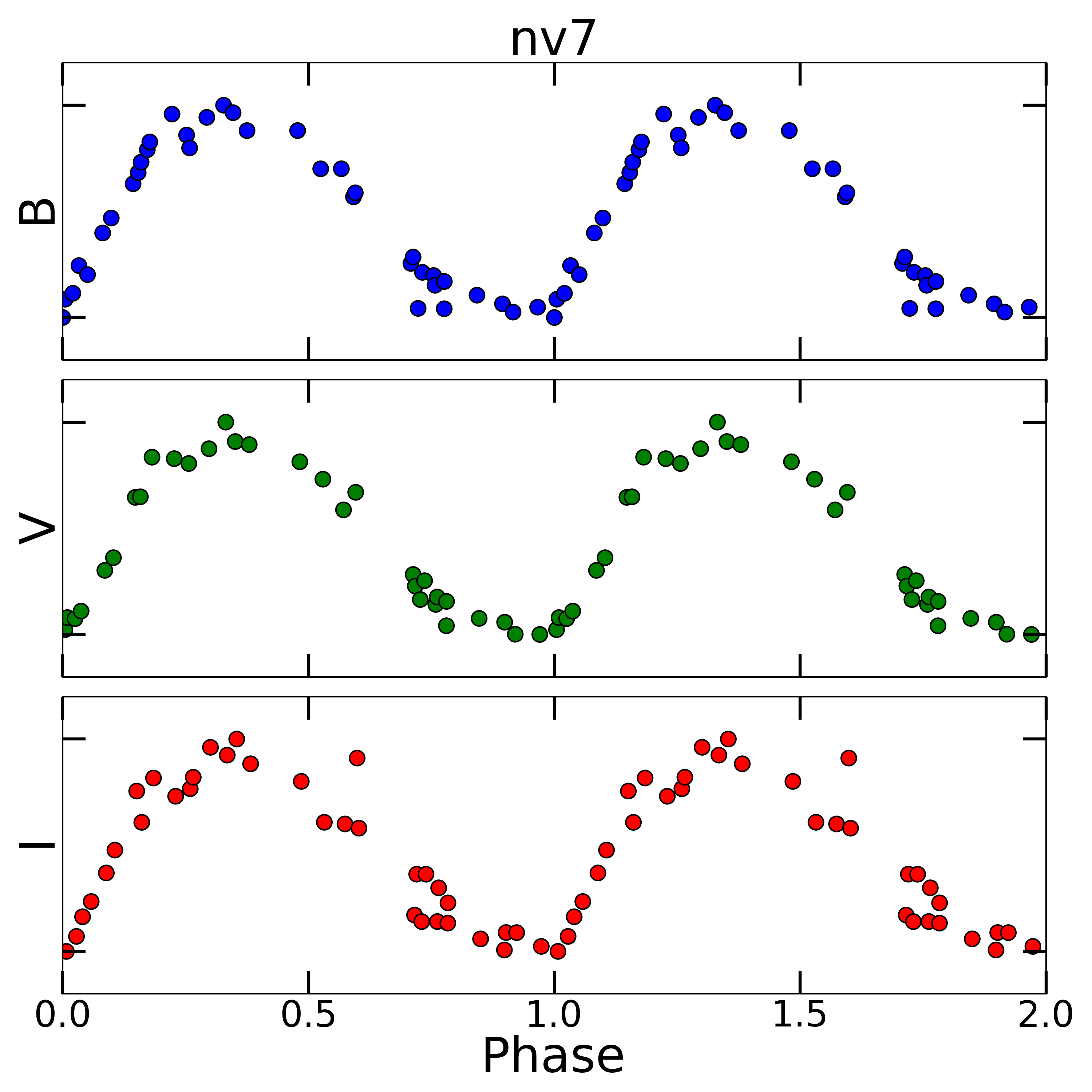}
   \caption{As in Figure~\ref{fig:rrab}, but for the c-type RR Lyrae.}
 \label{fig:rrc}
 \end{figure*}
}

\onlfig{4}{
 \begin{figure*}
   \centering
\includegraphics[width=.325\textwidth]{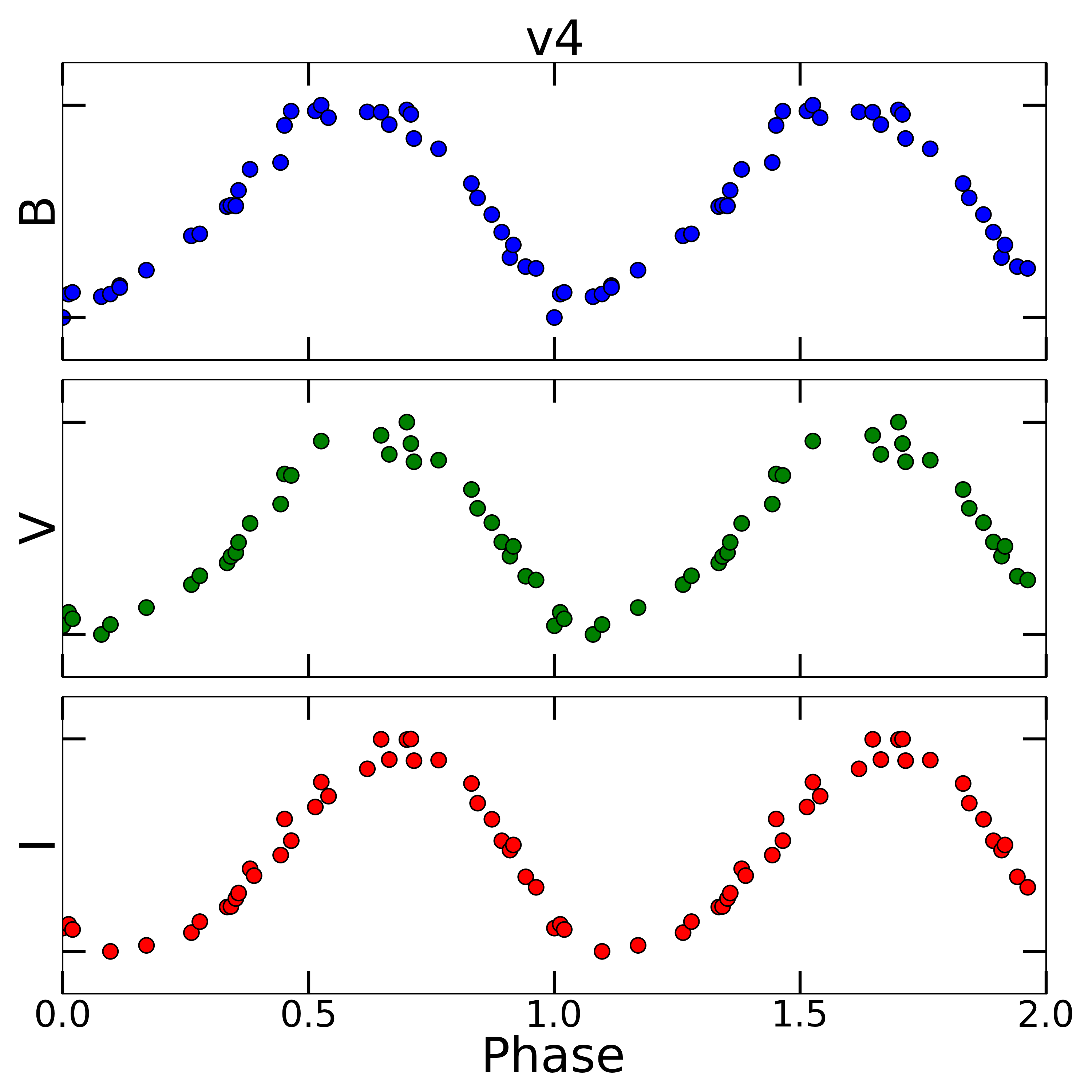}
\includegraphics[width=.325\textwidth]{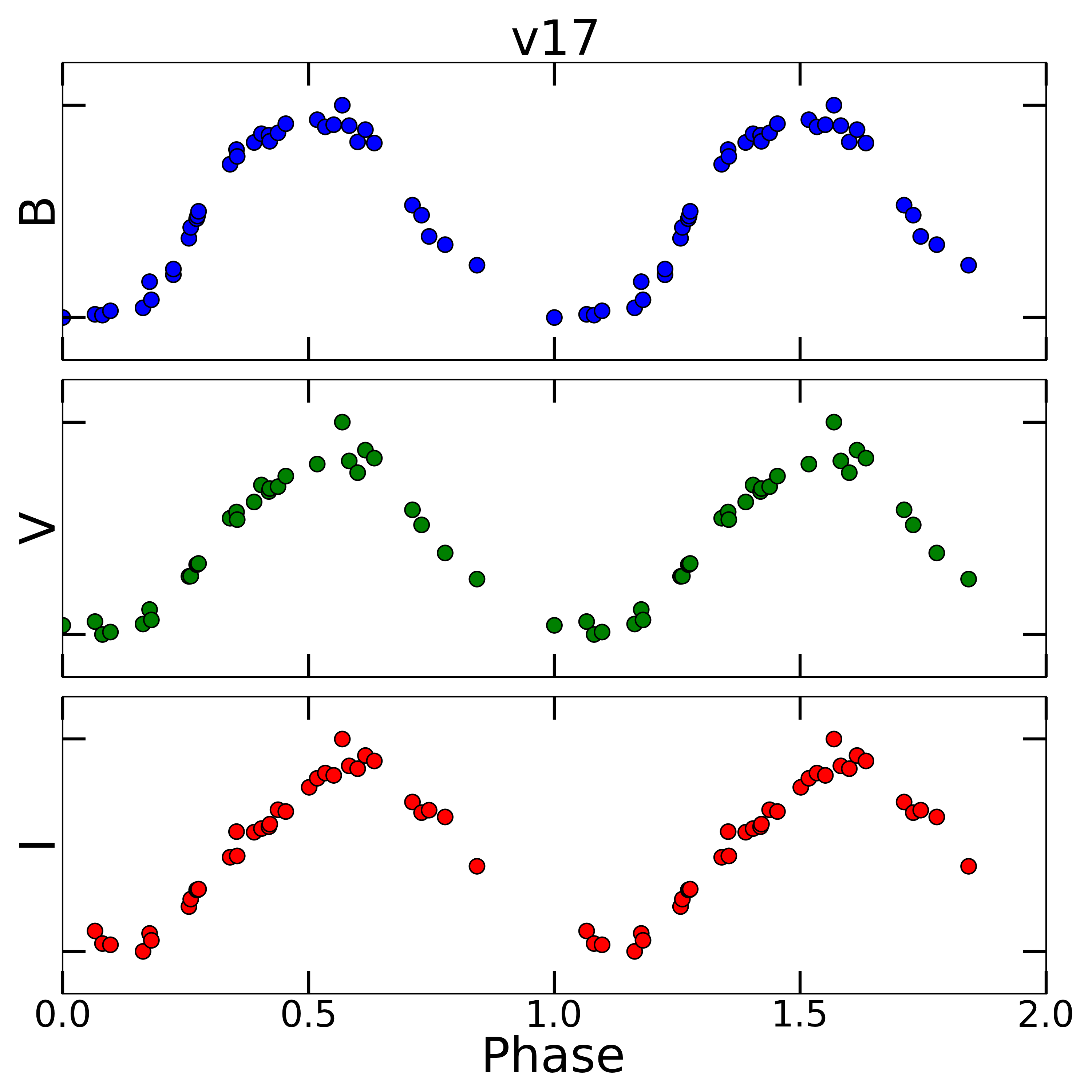}
   \caption{As in Figure~\ref{fig:rrab}, but for the type II Cepheids.}
 \label{fig:t2c}
 \end{figure*}
}

\onlfig{5}{
 \begin{figure*}
   \centering
\includegraphics[width=.325\textwidth]{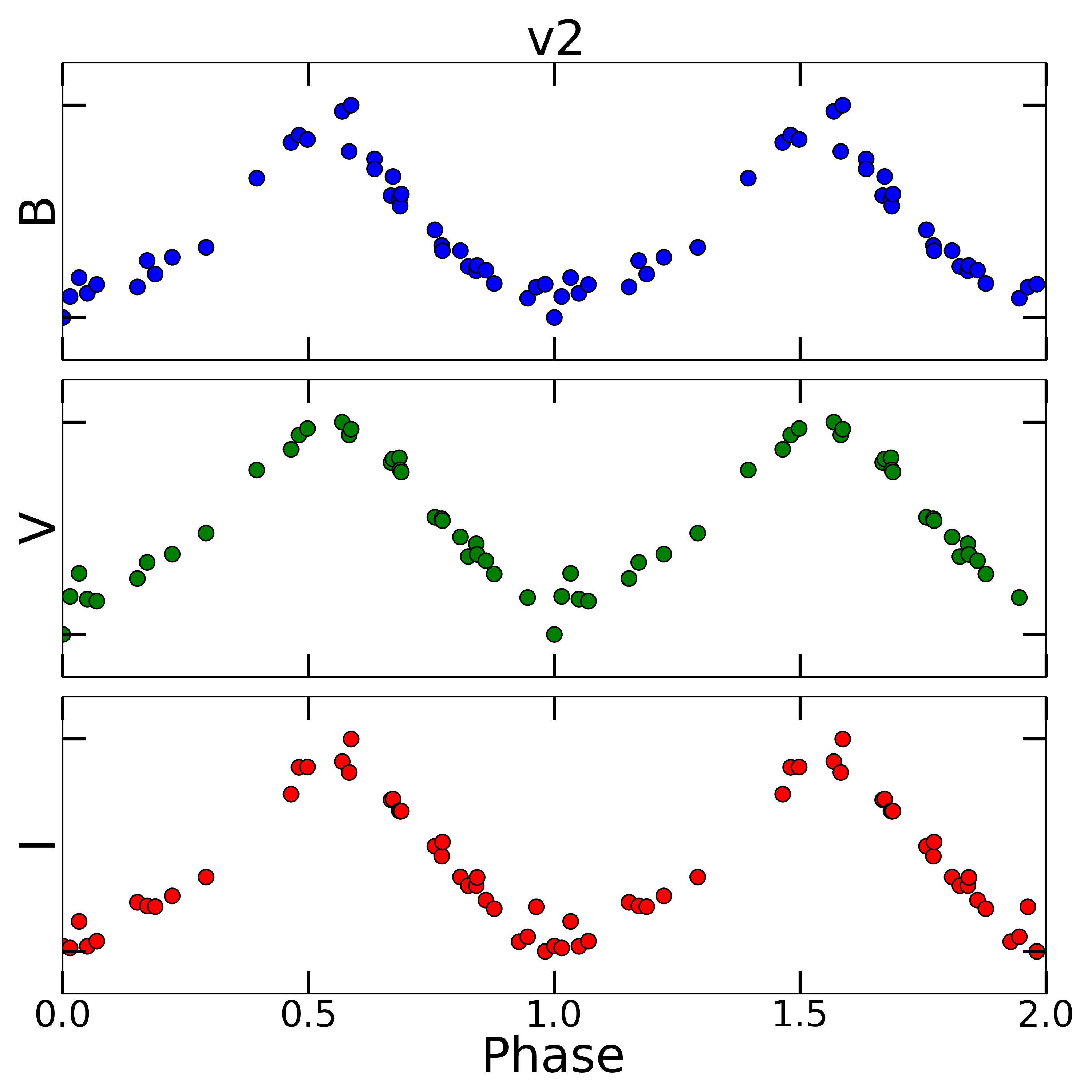}
\includegraphics[width=.325\textwidth]{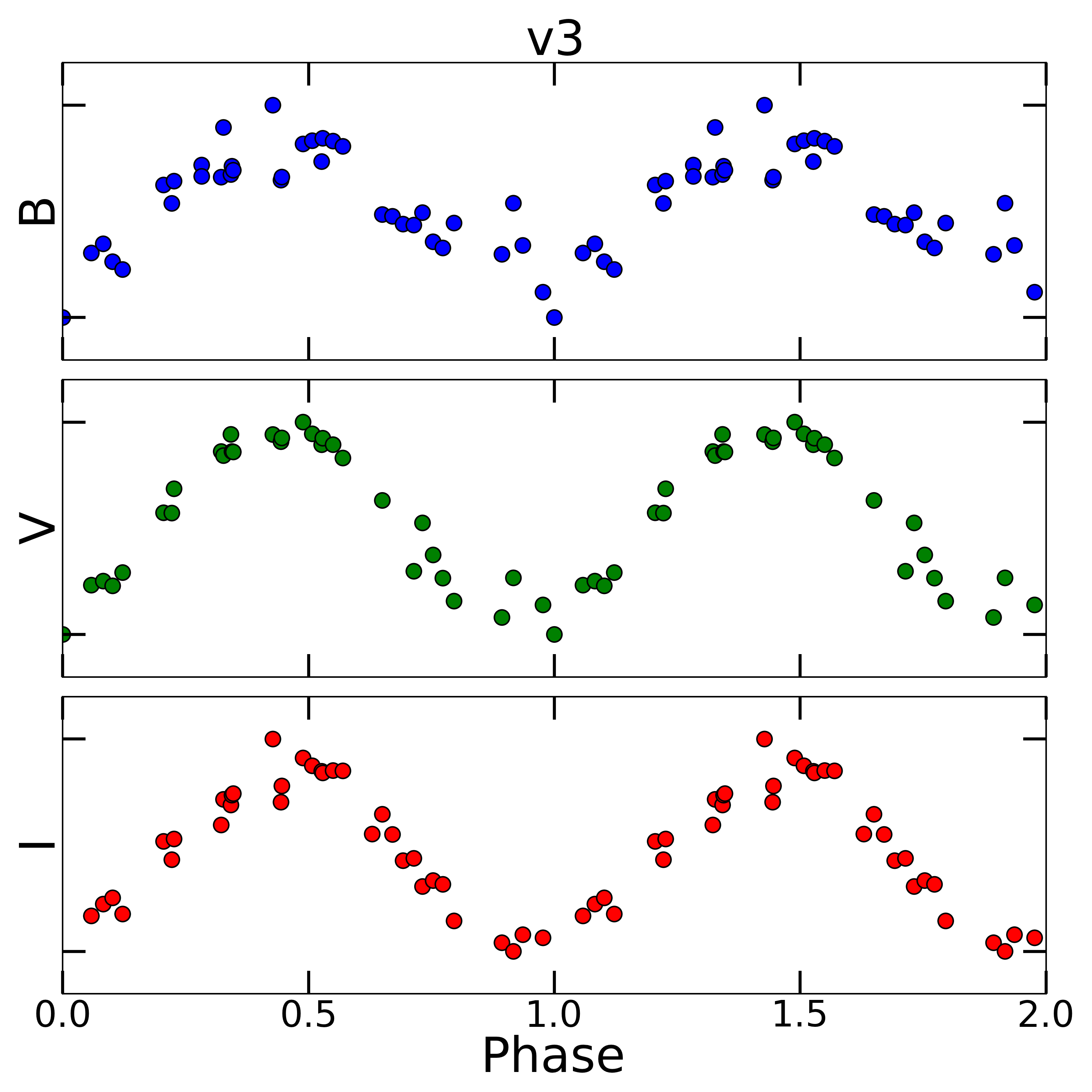}
\includegraphics[width=.325\textwidth]{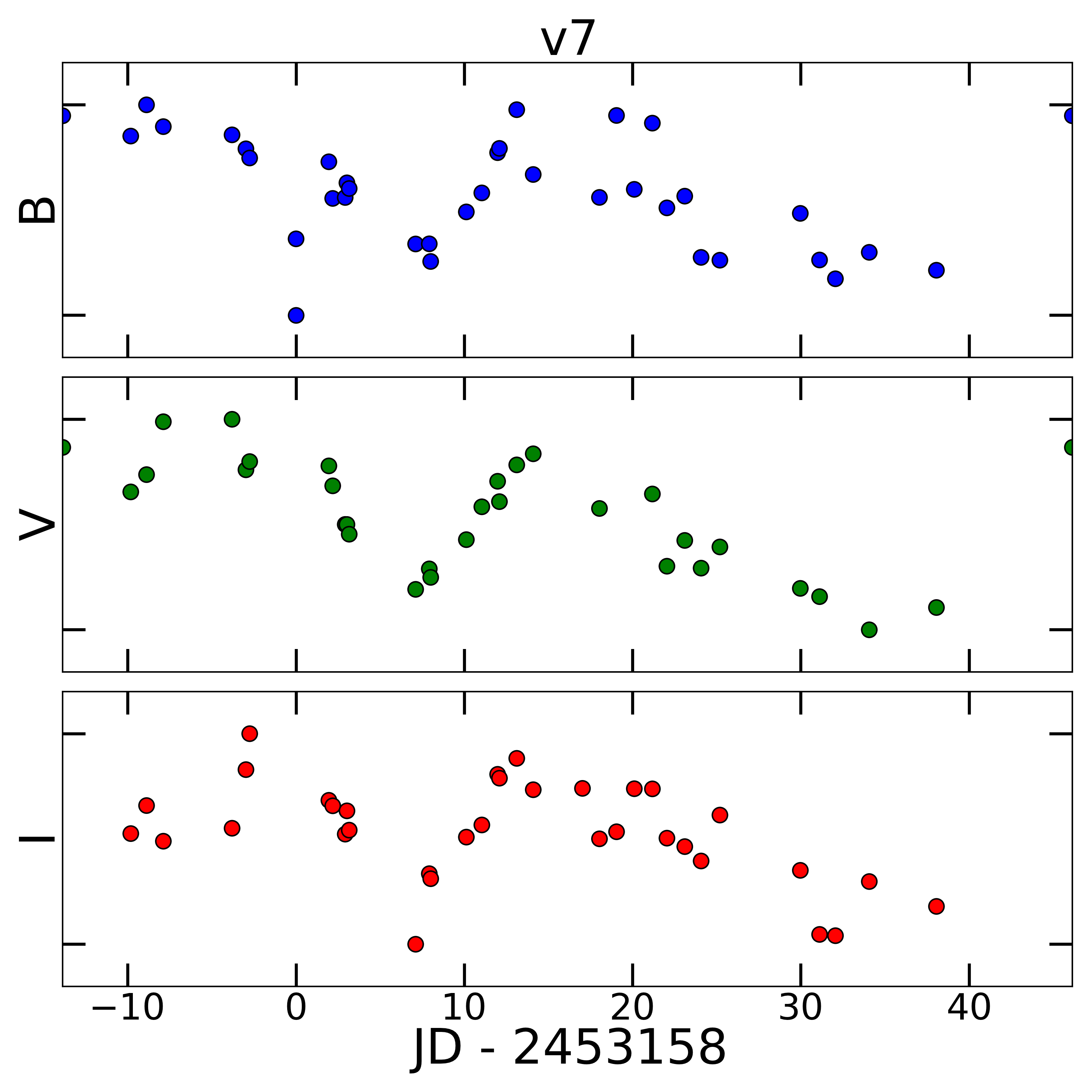}
\includegraphics[width=.325\textwidth]{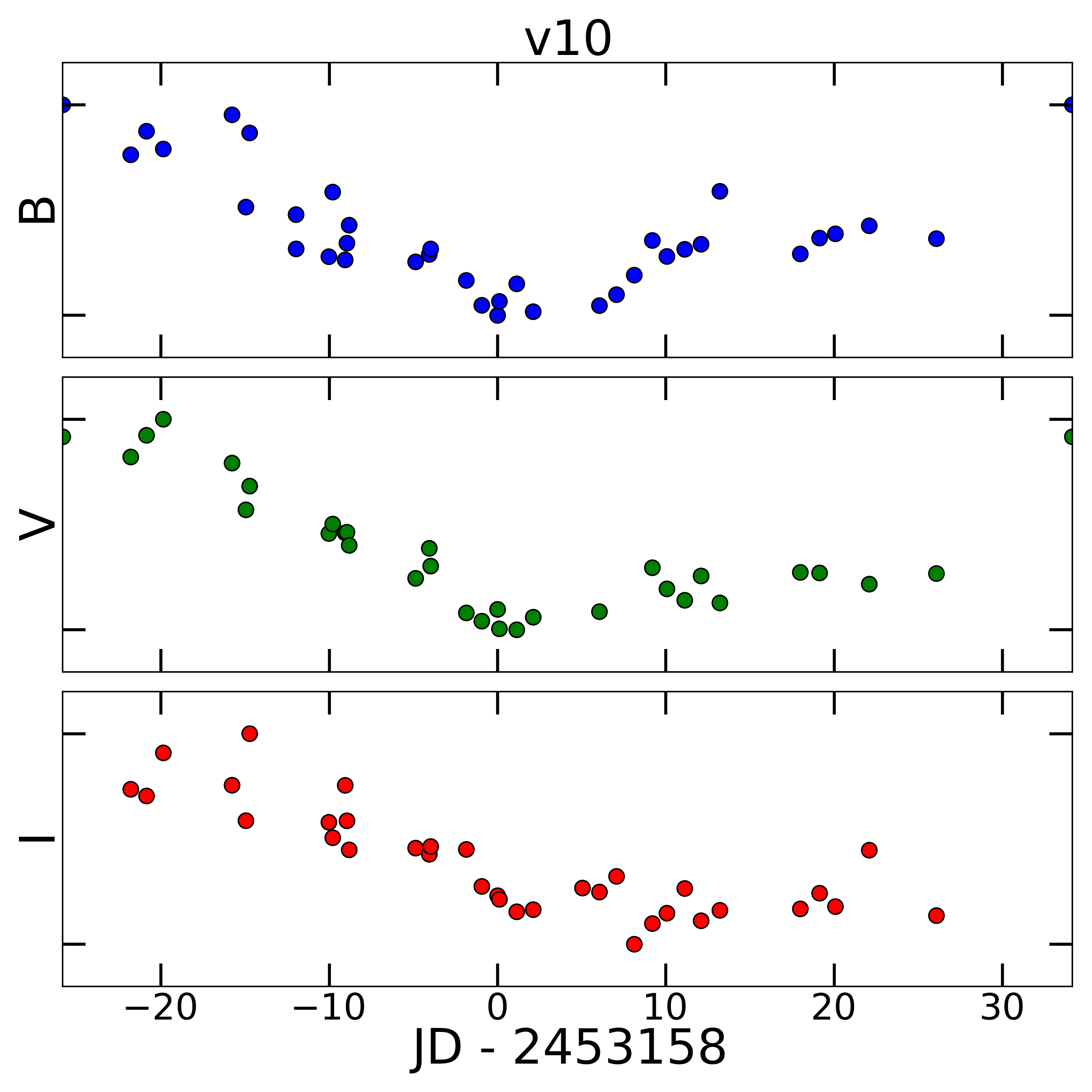}
\includegraphics[width=.325\textwidth]{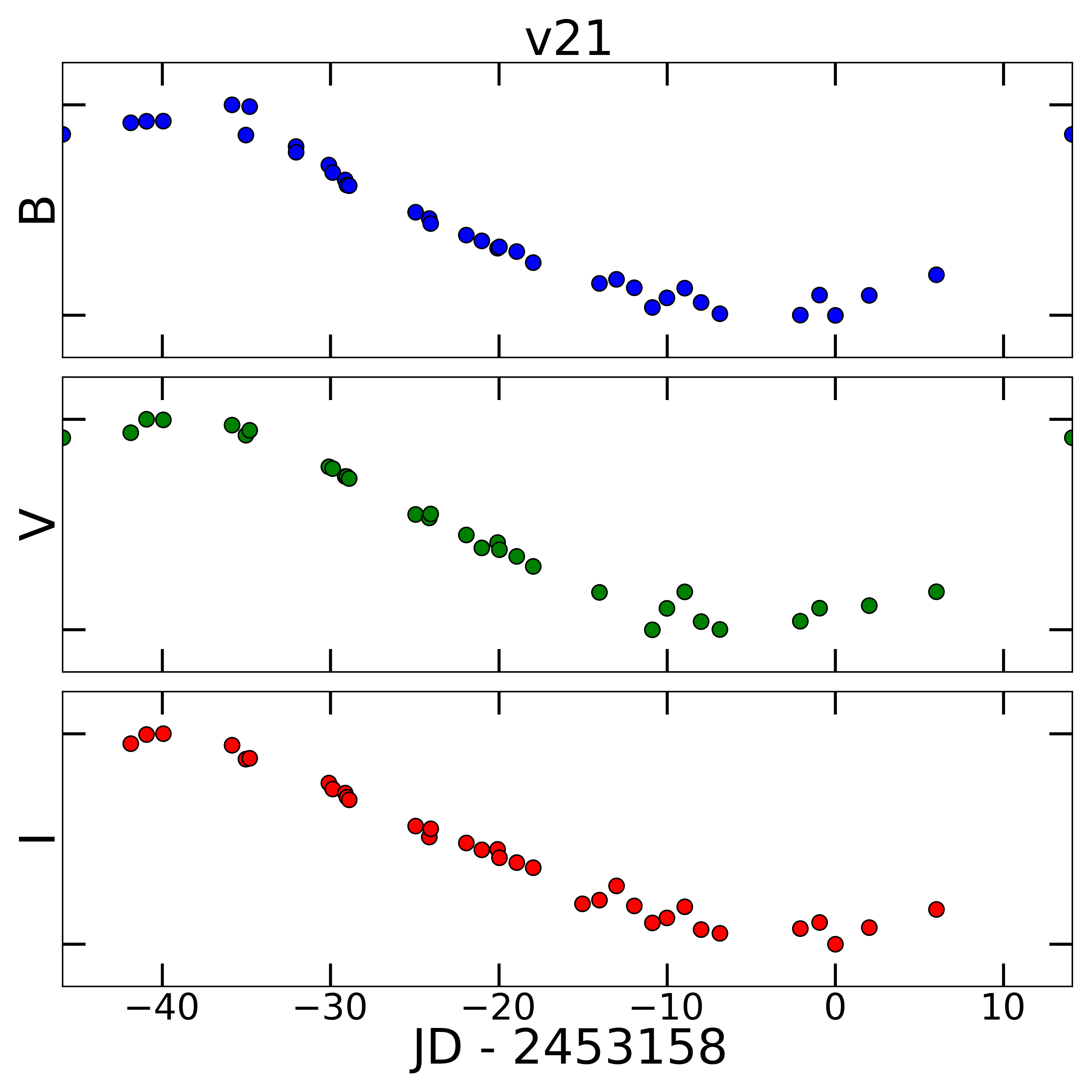}
\includegraphics[width=.325\textwidth]{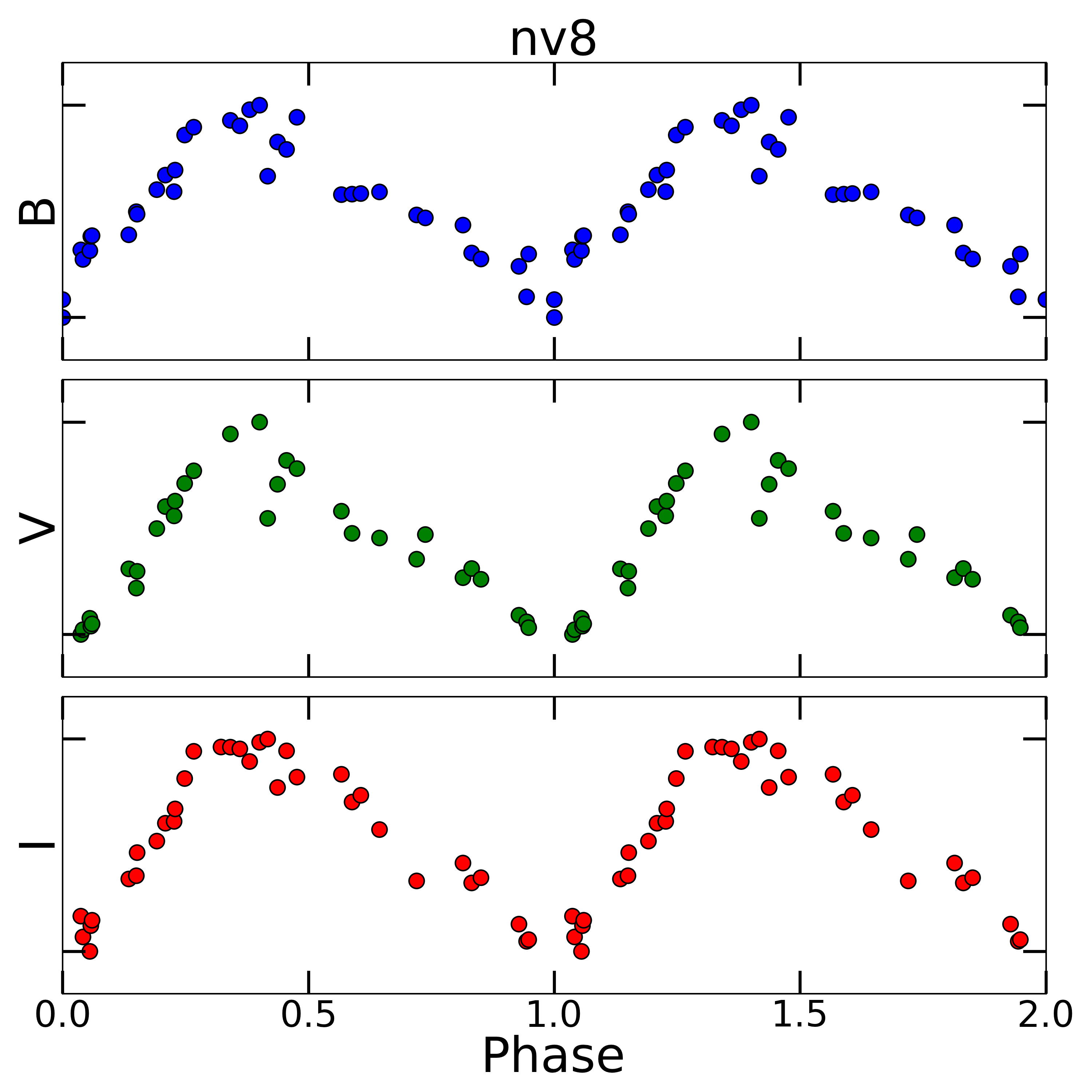}
\includegraphics[width=.325\textwidth]{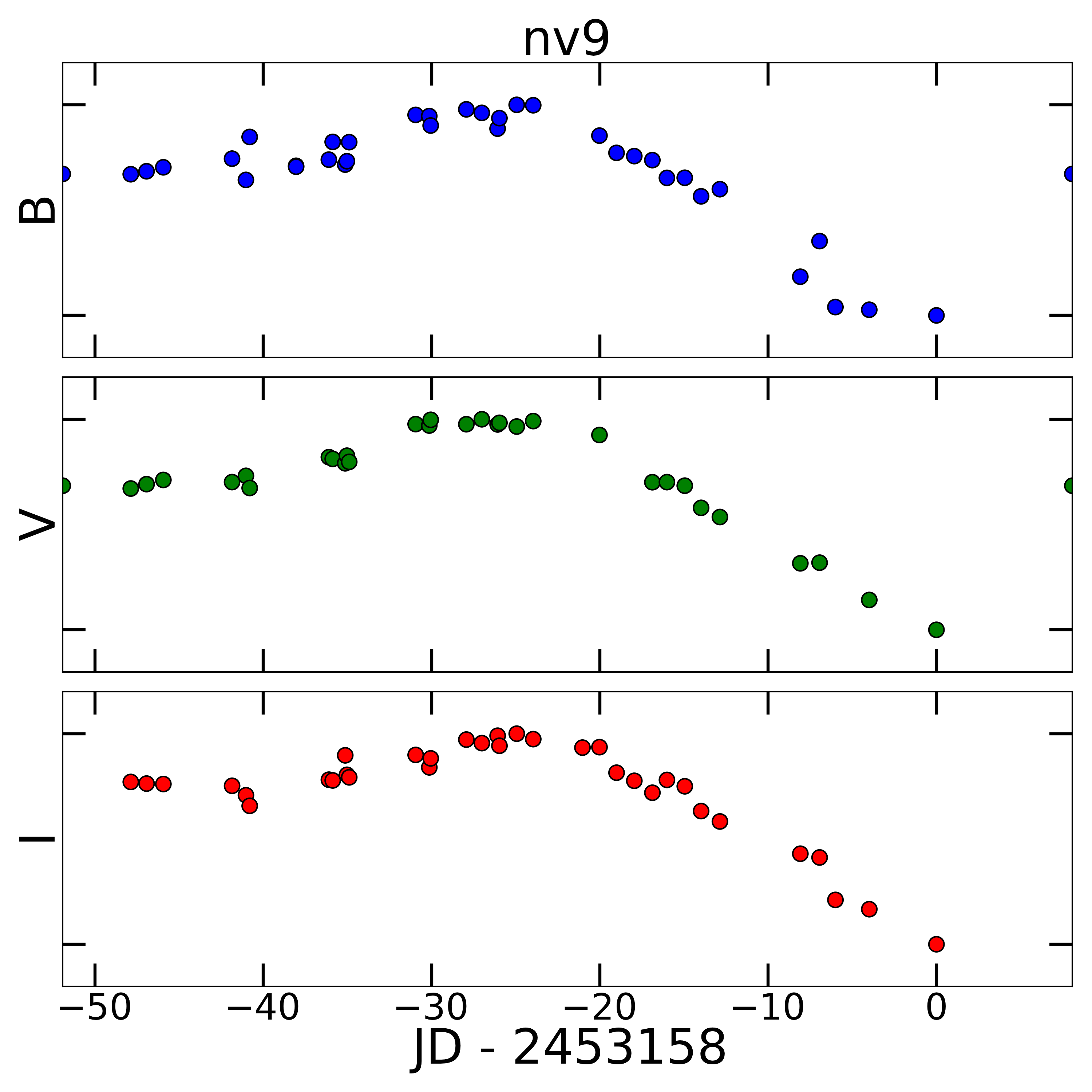}
   \caption{As in Figure~\ref{fig:rrab}, but for the LPVs.}
 \label{fig:lpv}
 \end{figure*}
}

\section{Summary}\label{sec:sum}
We have presented the results of a new search for variable stars in M28. Our search has led to the discovery of a number of previously unknown variables, most of which are c-type RR Lyrae stars. The properties of the ab-type RR Lyrae stars are most consistent with an Oosterhoff-intermediate classification, but this is not clearly supported by the properties of the c-type RR Lyrae stars. A ``hybrid'' Oosterhoff I/II classification is thus possible, raising the question as to whether multiple populations may be present in this fairly massive cluster. More extensive, higher-quality datasets will be required to put the properties of this cluster on a firmer basis.

\begin{acknowledgements}
We warmly thank C. M. Clement, E. D'Alessandro, J. Kaluzny, E. Tolstoy (the editor), and an anonymous referee for several comments that helped us improve our presentation. Support for M.C. and J.A.-G. is provided by the Ministry for the Economy, Development, and Tourism's Programa Inicativa Cient\'{i}fica Milenio through grant P07-021-F, awarded to The Milky Way Millennium Nucleus; by Proyecto Basal PFB-06/2007; by Proyecto FONDECYT Regular \#1110326; and by Proyecto Anillo ACT-86. H.A.S. thanks the US National Science Foundation for support under grants AST 0607249 and AST 0707756. This publication makes use of data products from the Two Micron All Sky Survey, which is a joint project of the University of Massachusetts and the Infrared Processing and Analysis Center/California Institute of Technology, funded by the National Aeronautics and Space Administration and the National Science Foundation.

\end{acknowledgements}

\end{document}